\documentclass[twocolumn,floatfix,superscriptaddress,pra]{revtex4-1}

\usepackage{times}
\usepackage{amssymb}
\usepackage{color,graphicx}
\usepackage{amsmath}
\usepackage{amsbsy}
\usepackage{amsthm}
\usepackage{bbm}
\usepackage{bm}
\usepackage{epsfig}
\usepackage{float}
\usepackage{natbib}

\definecolor{myurlcolor}{rgb}{0,0,0.7}
\definecolor{myrefcolor}{rgb}{0.8,0,0}
\usepackage[unicode=true,pdfusetitle, bookmarks=true,bookmarksnumbered=false,bookmarksopen=false, breaklinks=false,pdfborder={0 0 0},backref=false,colorlinks=true, linkcolor=myrefcolor,citecolor=myurlcolor,urlcolor=myurlcolor]{hyperref}

\newcommand{\dd}{\mathrm{d}}
\newcommand{\ket}[1]{\left| {#1} \right\rangle}
\newcommand{\bra}[1]{\left\langle {#1}\right|}
\newcommand{\ketbra}[2]{\ket{#1}\!\bra{#2}}
\newcommand{\braket}[2]{\langle #1|#2\rangle}

\newcommand{\mat}[2]{\left( \begin{array}{#1} #2  \end{array}\right)}

\newcommand{\ii}{\mathrm{i}}
\newcommand{\ee}{\mathrm{e}}
\renewcommand{\t}[1]{\textrm{#1}}

\newcommand{\ot}[0]{\otimes}

\newcommand{\para}[0]{\parallel}

\newcommand{\eref}[1]{(\ref{#1})}
\newcommand{\eqnref}[1]{Eq.~(\ref{#1})}
\newcommand{\eqnsref}[2]{Eqs.~(\ref{#1}) and (\ref{#2})}
\newcommand{\figref}[1]{Fig.~\ref{#1}}

\newcommand{\appref}[1]{App.~\ref{#1}}

\begin{document}
\title{The ultimate precision limits for noisy frequency estimation}

\author{Andrea Smirne}
\affiliation{Institute of Theoretical Physics, Universit{\"a}t Ulm, Albert-Einstein-Allee 11D-89069 Ulm, Germany}

\author{Jan Ko\l{}ody\'{n}ski}
\affiliation{ICFO-Institut de Ci\`encies Fot\`oniques, Mediterranean Technology Park, 08860 Castelldefels (Barcelona), Spain}

\author{Susana F. Huelga}
\affiliation{Institute of Theoretical Physics, Universit{\"a}t Ulm, Albert-Einstein-Allee 11D-89069 Ulm, Germany}

\author{Rafa\l{} Demkowicz-Dobrza\'{n}ski}
\affiliation{Faculty of Physics, University of Warsaw, 02-093 Warszawa, Poland}

\begin{abstract}
Quantum metrology protocols allow to surpass precision limits typical to classical statistics.
However, in recent years, no-go theorems have been formulated,
which state that typical forms of uncorrelated noise can constrain the quantum enhancement to a constant
factor, and thus bound the error to the standard asymptotic scaling. 
In particular, that is the case of time-homogeneous (Lindbladian) dephasing and, more generally,
all semigroup dynamics that include \emph{phase covariant} terms, which commute with the system Hamiltonian.
We show that the standard scaling can be surpassed when the dynamics is no longer ruled by a semigroup 
and becomes time-inhomogeneous. In this case, the ultimate precision is determined by the system
short-time behaviour, which when exhibiting the natural Zeno regime leads to
a non-standard asymptotic resolution. In particular, we demonstrate that the relevant noise
feature dictating the precision is the violation of the semigroup property at
short timescales, while non-Markovianity does not play any specific role.
\end{abstract}

\maketitle

\textit{Introduction.---}%
Parameter estimation, ranging from the precise determination of atomic transition frequencies to external 
magnetic field strengths, is a central task in modern physics \cite{Caves1981,Wineland1992,Giovannetti2004,Leibfried2004,Roos2006,Pezze2009,Jones2009}.
Quantum probes made up of $N$ entangled particles can attain the so-called \emph{Heisenberg limit} (HL),
where the estimation \emph{mean squared error} (MSE) scales as $\sim\!1/N^2$, as compared with 
the \emph{standard quantum limit} (SQL) $\sim\!1/N$ of classical statistics 
\cite{Demkowicz2015,*Toth2014}.

Heisenberg resolution relies on the unitarity of the time evolution. In realistic situations, however, quantum 
probes decohere as a result of the unavoidable interaction with the surrounding environment
\cite{Breuer2002,*Rivas2012}. Such interactions can have a dramatic effect on estimation precision---even
infinitesimally small uncorrelated dephasing noise, modelled as a semigroup (time-homogeneous-Lindbladian) 
evolution \cite{Gorini1976,*Lindblad1976}, forces the MSE to eventually follow the SQL \cite{Huelga1997}. 
This result was proven to be an instance of the \emph{quantum Cram\'{e}r-Rao bound} (QCRB) \cite{Braunstein1994} 
for generic Lindbladian dephasing and thus holds even when using optimized entangled states and measurements \cite{Escher2011,Demkowicz2012,Kolodynski2013,*Kolodynski2014,Knysh2014}.
The question then arises of what is the ultimate precision limit when the noisy time evolution is not governed 
by a dephasing dynamical semigroup
\cite{
Escher2011,
Demkowicz2012,
Kolodynski2013,*Kolodynski2014,
Knysh2014,
Ulam2001,Shaji2007,Fujiwara2008,Huver2008,
Dorner2009,*Knysh2011,*Kacprowicz2010,*Kolodynski2010,
Brivio2010,*Genoni2011,
Chwedenczuk2012,
Matsuzaki2011,*Chin2012,
Chaves2013,*Brask2015,
Berrada2013}. 
The SQL has been shown to be surpassable in the presence of time-inhomogeneous (non-semigroup) dephasing 
noise \cite{Matsuzaki2011,*Chin2012}, noise with a particular geometry \cite{Chaves2013,*Brask2015}
and correlated time-homogeneous dephasing \cite{Dorner2012,*Jeske2014}, or
when the noise geometry allows for error correction techniques 
\cite{Kessler2014,*Duer2014,*Arrad2014,*Lu2015,*Plenio2015}.

\begin{figure}[!t]
\includegraphics[width=.85\columnwidth]{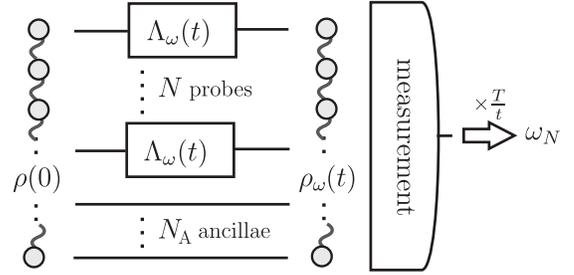}
\caption{\textbf{Noisy fequency estimation scenario.}~$N$
qubit probes sense a parameter $\omega$ following a preparation in a state $\rho(0)$, including
an arbitrary number $N_\t{A}$ of ancillary particles potentially
entangled with the sensing probes. During the evolution, the probes are
subject to uncorrelated noise and after time $t$ the whole system, in state $\rho_\omega(t)$,
is measured. The protocol is repeated $T/t$ times, with $T\!\gg\!t$,
to construct a frequency estimate $\omega_N$.
}
\label{fig:setup}
\end{figure}
Here, we derive the ultimate lower bounds on the MSE
for the \emph{noisy frequency estimation} scenario depicted in \figref{fig:setup}
where probe systems are independently affected by the decoherence. In particular, we
focus on uncorrelated \emph{phase-covariant} noise, that is, noise-types commuting with 
the parameter-encoding Hamiltonian, as these underpin the asymptotic SQL-like precision in 
the semigroup case \citep{Chaves2013,Knysh2014}. Yet, most importantly,
we allow for any form of time-inhomogeneity and non-Markovian features in the noise. 
Our results show that, when moving away from the
semigroup regime, entanglement generally improves the precision \emph{beyond}
the constant-factor enhancement, so that the SQL is truly overcome.
As a special case, we confirm the conjecture made in \cite{Matsuzaki2011,*Chin2012},
where by considering a Ramsey interferometry scheme and
non-semigroup dephasing dynamics, a $1/N^{3/2}$ error scaling was shown to be achievable.
This was argued to be a consequence of the \emph{Zeno regime} at short time scales. The 
generality of this scaling has been recently verified for pure dephasing noise \cite{Macieszczak2015}.
We formally prove the emergence of non-SQL scaling for any non-semigroup phase 
covariant noise. We demonstrate that it is \emph{solely} the short-time expansion of the 
effective noise parameters that determines the ultimate attainable precision. In particular, any memory
(non-Markovian) effects, which may be displayed by the system at later times,
are irrelevant for the asymptotic $N$ limit.

\textit{Noisy frequency estimation.---}%
In a typical frequency estimation setting,
a parameter $\omega$ is unitarily encoded on $N$ sensing particles
(probes), specifically qubits, over the interrogation time $t$ during which the probes are also
independently disturbed by the decoherence \cite{Huelga1997,Escher2011}.
As depicted in \figref{fig:setup}, we generalise such a setup to allow
for an arbitrary number $N_\t{A}$ of ancillary particles, that can be initially entangled
with the probes and measured at the end of the protocol.
Hence, the combined final state of the system reads:
\begin{equation}
\rho_\omega(t) = \Lambda_{\omega}(t)^{\ot N} \ot \mathbbm{1}^{\ot N_\t{A}} [\rho(0)],
\label{eq:fin_state}
\end{equation}
with $\rho(0)$ being the initial state, and $\Lambda_{\omega}(t)$
a \emph{completely positive and trace preserving} (CPTP) linear map
\cite{Bengtsson2006} representing the identical, but independent,
evolution of each probe (see \appref{app:cptp_maps_reps}).
We assume full control and noise-free evolution for the ancillae, so that to 
allow for single-step error-correction protocols \cite{Kessler2014,*Duer2014,*Arrad2014,*Lu2015}.
The $N$ dependent parameter estimate, $\omega_N$, relies on sufficiently large
statistical data after performing $T/t$ repetitions, provided the 
total experimental time $T\!\gg\!t$.

We quantify the performance of the estimation protocol by the MSE, $\Delta^2 \omega_N$---%
describing the average deviation of the estimate from the true value. Crucially,
requiring unbiasedness and consistency for the estimate, the
QCRB directly provides us with the ultimate lower bound on the MSE 
that is optimised over all potential measurement strategies \cite{Braunstein1994}.
Hence, possessing also the freedom to adjust the single-shot duration time $t$, the ultimate attainable precision can be written as
\begin{equation}
\label{eq:una}
\Delta^2 \omega_N\, T \geq \min_{t} \frac{t}{F_\t{Q}[\rho_{\omega}(t)]},
\end{equation}
where $F_\t{Q}[\rho_{\omega}(t)]$ is the \emph{quantum Fisher Information} (QFI)
evaluated with respect to (w.r.t.)~the estimated parameter $\omega$ encoded in the final state \eref{eq:fin_state}.
Importantly, the $t$ that minimises the right-hand side in \eqnref{eq:una}, i.e., the \emph{optimal single-shot
duration}, generally depends on the system size and we thus denote it as $t_\t{opt}(N)$.

\begin{figure}[!t]
\includegraphics[width=.75\columnwidth]{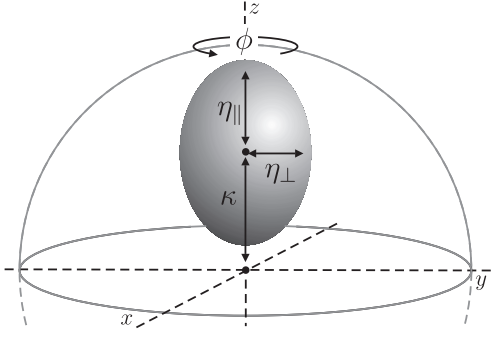}
\caption{\textbf{Phase covariant quantum maps.} Bloch ball representation of a
qubit noisy evolution, $\Lambda_\omega(t)$, that commutes with the rotation
about the $z$ axis by an angle $\omega t$, which represents the parameter encoding.
The overall effect of the noise is to shrink the ball by factors $\eta_\parallel(t)$, $\eta_\perp(t)$
in the vertical direction and the horizonal plane respectively,
as well as to displace its the centre by $\kappa(t)$ and to further rotate it about the $z$ axis, so that the global rotation angle is
$\phi\!=\!\omega t + \theta$.
}
\label{fig:cov_maps}
\end{figure}

\textit{Phase covariant dynamics.---}%
The frequency parameter $\omega$ is unitarily encoded within the phase, $\omega t$,
accumulated during the free evolution of qubit probe, which in the Bloch ball picture
corresponds to a rotation around a known direction---$z$ in \figref{fig:cov_maps}.
We consider systems exhibiting uncorrelated forms of noise that
commute with such rotations, which formally correspond to the so-called
\emph{phase covariant} qubit maps \cite{Holevo1993,*Holevo1996}.
Such noise-types are known to most severely limit the attainable precision
in case of semigroup dynamics, for which they constrain the quantum
enhancement to a constant factor above the SQL \cite{Escher2011,Demkowicz2012,Kolodynski2013,Kolodynski2014,Knysh2014}.
Although such negative conclusion cannot be drawn for other
less severe but still semigroup noises:~purely transversal \cite{Chaves2013}
and correlated \cite{Dorner2012,*Jeske2014}; the
phase covariant noise if present, no matter how weak, must always
asymptotically dominate and limit the ultimate quantum improvement
to a constant factor \cite{Chaves2013,Knysh2014}. Thus, in what follows, we focus on the
frequency estimation scenario of \figref{fig:setup} in the presence of general
\emph{independent, identical and phase covariant} (IIC) noise,
where each single probe at any instance of time may be described by the action of a map
\begin{equation}
\Lambda_\omega(t) = \mathcal{U}_{\omega}(t) \circ \Gamma(t) = \Gamma(t) \circ \mathcal{U}_{\omega}(t)
\label{eq:qubit_evol}
\end{equation}
with $\mathcal{U}_\omega(t)$ and $\Gamma(t)$ being its unitary encoding and $\omega$-independent dissipative
parts respectively. We fix 
$\mathcal{U}_{\omega}(t)[\varrho]\!=\!\ee^{-\frac{\ii}{2} \omega \sigma_z t} \varrho \ee^{\frac{\ii}{2} \omega \sigma_z t}$
and, as shown in \appref{app:cov_maps_reps}, use the affine representation of qubit maps
\cite{King2001, *Ruskai2002,Andersson2007,
Asorey2009, Chruscinski2010b,*Chruscinski2010, Smirne2010}
to express the most general phase covariant $\Lambda_\omega(t)$ as a matrix:
 \begin{equation}
\label{eq:phasecov_map}
{\sf\Lambda}_{\omega}(t)= \mat{cccc}{
1 & 0 & 0 & 0\\
0 & \eta_{\perp}(t) \,\cos\phi(t)  &  -\eta_{\perp}(t) \,\sin\phi(t) & 0 \\
0 & \eta_{\perp}(t)\,\sin\phi(t)  & \eta_{\perp}(t) \,\cos \phi(t)  & 0 \\
\kappa(t) & 0 & 0 & \eta_{\parallel}(t)}\!,
\end{equation}
that acts on a four-component Bloch vector.
As depicted in \figref{fig:cov_maps}, the qubit evolution amounts then to:~a
rotation around the $z$ axis by an angle $\phi$ containing the parameter encoding
($\phi(t) \!=\! \omega t \!+\! \theta(t)$), a contraction in the $xy$ plane by a factor $0\!\leq\! \eta_\perp \!\leq\! 1$,
a contraction in the $z$ direction by a factor $0 \!\leq\! |\eta_\parallel| \!\leq\! 1$
($\eta_\parallel\!<\!0$ case corresponds to an additional reflection with respect to the $xy$ plane),
and a displacement in the z direction by $-1 \!\leq\! \kappa \!\leq\! 1$.
The map \eref{eq:phasecov_map} fulfils the CPTP condition as long as
$\eta_\parallel \!\pm\! \kappa \!\leq\!1$ and $1+\eta_\parallel \!\geq\! \sqrt{4 \eta_{\perp}^2 \!+\! \kappa^2}$
(see \appref{app:cov_maps_reps}).

It is important to stress that any single qubit phase covariant dynamics $\varrho(t) \!=\! \Lambda_{\omega}(t)[\varrho(0)]$ can always be put on physical grounds
by considering a corresponding time-local master equation of the form
\begin{eqnarray}
\frac{\dd}{\dd t} \varrho(t) &=& -\frac{\ii}{2} (\omega + h(t)) [\sigma_z,  \varrho(t)] \nonumber\\
&&+ \gamma_+(t)\left(\sigma_+ \varrho(t) \sigma_- - \frac{1}{2} \left\{\sigma_- \sigma_+, \varrho(t)\right\}\right) \nonumber \\
&& + \gamma_-(t)\left(\sigma_- \varrho(t) \sigma_+ - \frac{1}{2} \left\{\sigma_+ \sigma_-, \varrho(t)\right\}\right) \nonumber\\
&&+ \gamma_z(t) \left(\sigma_z \varrho(t) \sigma_z -  \varrho(t) \right).
\label{eq:step1me}
\end{eqnarray}
The proof as well as explicit relations between $\theta$, $\eta_\perp$, $\eta_\parallel$, $\kappa$ and
$h$, $\gamma_+$, $\gamma_-$, $\gamma_z$ are given in \appref{app:cov_maps_reps}.
Phase covariant dynamics therefore describes any physical evolution
that may arise from combinations of time-varying absorption, emission and dephasing processes,
as well as Lamb shift corrections to free Hamiltonian \cite{Breuer2002,*Rivas2012}.
Moreover, \eqnref{eq:step1me} generally allows for the quantitative characterisation of non-Markovian effects 
\cite{Breuer2009,*Breuer2015,Rivas2010,*Rivas2014}.
In the special case of positive constant rates (time homogeneity), \eqnref{eq:step1me} provides the generator
of any phase covariant quantum dynamical semigroup \cite{Vacchini2010}.

\textit{Bounding the ultimate precision.---}%
Having fully characterized the class of qubit IIC dynamics, we can now state the main result of the paper.
Given $N$ qubit probes and $N_A$ ancillae evolving according to \eqnref{eq:fin_state}, with the single qubit dynamics
given by a phase covariant map $\Lambda_\omega(t)$ as in \eqnref{eq:phasecov_map},
and provided that at all times ($\forall t\!>\!0$) $\eta_{\perp}(t) \!<\!1$, the MSE in estimating the frequency
$\omega$ is asymptotically determined by the short-time expansion of the noise parameters:
\begin{eqnarray}
&\eta_{\perp}(t)=
1 - \alpha_\perp t^{\beta_\perp} + o(t^{\beta_\parallel}),
\quad
\kappa(t) = \alpha_\kappa t^{\beta_\kappa}  +  o(t^{\beta_\kappa})& \nonumber \\
&
\eta_{\parallel}(t) =1  -  \alpha_\parallel t^{\beta_\parallel} +  o(t^{\beta_\parallel}),&
\label{eq:cov_coeffs}
\end{eqnarray}
and it satisfies the following inequality
\begin{equation}
\lim_{N \rightarrow \infty} \frac{\Delta^2 \omega_N\,T}{N^{-(2 \beta_\perp-1)/\beta_\perp}}
\;\geq\;
\frac{\alpha^{1/\beta_{\perp}} \beta_{\perp}}{(\beta_{\perp}-1)^{(\beta_{\perp}-1)/\beta_{\perp}}}
=D,
\label{eq:step2}
\end{equation}
where $D\!>\!0$ and
\begin{equation}
\alpha =
\left\{
\begin{array}{cc}
2 \alpha_{\perp} &\quad \beta_{\perp}< \beta_{\parallel}; \\
2 \alpha_{\perp}-\frac{\alpha_{\parallel}}{2} &\quad  \beta_{\perp}= \beta_{\parallel}<\beta_\kappa; \\
\max\left\{2 \alpha_{\perp}-\frac{\alpha_{\parallel}}{2} -\frac{|\alpha_{\kappa}|}{2}, \frac{|\alpha_{\kappa}|}{4}\right\}&\quad  \beta_{\perp}= \beta_{\parallel}= \beta_\kappa.
\end{array}\right.
\label{eq:fincases2}
\end{equation}

Crucially, as (see below) the bound in \eqnref{eq:step2} is always attainable up to
a constant factor,
the asymptotic precision is fully determined by the short-time expansion of the
radius in the plane perpendicular to the rotation axis, which
fixes the asymptotic scaling to $1/N^{(2\beta_\perp-1)/\beta_\perp}$.
For semigroup dynamics ($\beta_\perp\!=\!1$) one accordingly recovers the
SQL-like $1/N$ limit, while with increasing $\beta_\perp$ one finds a progressively more favourable
scaling that tends to HL for unrealistic $\beta_\perp\!\rightarrow\!\infty$.
Besides the assumption of IIC noise \eref{eq:qubit_evol}, the only condition assuring the
bound \eref{eq:step2} to be valid is $\eta_{\perp}(t)\!<\!1$. In fact, if $\eta_{\perp}(t)\!=\!1$ at some finite $t$
then by CPTP-property also $\eta_{\parallel}(t)\!=\!1$ and $\kappa(t)\!=\!0$. In other words, 
a  ``\emph{full revival}'' of the Bloch vector length occurs and the only effect of the 
interaction with the environment is a rotation about the $z$-axis by some angle $\theta$.
Not surprisingly, the best estimation strategy is then to measure the frequency
at such pseudo-noiseless moment, at which the $1/N^2$ HL is attainable.
However, note that such a behaviour is quite unlikely when
dealing with open systems subject to realistic sources of noise \cite{Breuer2002}.

The sketch of the proof is provided below, while a complete version
is given in \appref{app:proof}. Firstly, we fix the evolution time $t$ (and omit it for simplicity), to
use the finite-$N$ channel extension (CE) method \cite{Fujiwara2008,Kolodynski2013,*Kolodynski2014},
which provides an upper-bound on the QFI that is already optimised over all initial states:
\begin{multline}
\max_{\rho(0)} F_\t{Q}\!\left[\Lambda_{\omega}^{\ot N} \!\ot\! \mathbbm{1}^{\ot N_\t{A}}[\rho(0)]\right] \;\leq  \\
\leq\;4  N \min_{\{K_i\}}\{\| A \| + (N-1) \| B\|^2)\}= F^{\uparrow}.
\label{eq:CE_bound}
\end{multline}
The minimisation above is performed over Kraus representations of the
channel $\Lambda_{\omega}[\varrho]\!=\!\sum_iK_i\varrho K_i^\dagger$,
describing the dynamics of a \emph{single} probe.
$\|\!\cdot\!\|$ denotes the operator norm, whereas $A \!=\! \sum_{i} \dot{K}^\dagger_i \dot{K}_i$ and $B\!=\! \sum_i \dot{K}^\dagger_i K_i$
with $\dot{K}_i\!\equiv\!\tfrac{d}{d \omega} K_i$.
Identifying the optimal Kraus representation is usually non-trivial, however, the numerical
semidefinite programming (SDP) methods introduced in \cite{Kolodynski2013,*Kolodynski2014}
automatically provide the correct ansatz, with which one may then proceed analytically.

In \appref{app:proof}, we explicitly deal with the general case of phase covariant qubit map, where we additionally prove
the convexity of the bound \eref{eq:CE_bound} w.r.t.~mixing of quantum channels. This allows us to analytically apply
\eqnref{eq:CE_bound} to \emph{any} map $\Lambda_{\omega}$ of the form \eref{eq:phasecov_map}, after 
adequately decomposing it into an optimal mixture of unital ($\kappa\!=\!0$) and amplitude damping channels
($\eta_\para\!=\!\eta_\perp^2\!=\!1\!-\!\kappa$). Here, for simplicity, we focus on unital channels
for which $\kappa\!=\!0$ and the general upper bound \eref{eq:CE_bound} derived in \appref{app:proof} 
reduces to:
\begin{equation}
F^{\uparrow}_{\eta_\parallel,\eta_\perp} \!= \frac{t^2 N^2}{1+N \ell(t)},\quad \ell(t) = \frac{1+\eta_\parallel(t)-2\eta_\perp(t)^2}{2 \eta_\perp(t)^2}.
\label{eq:l_def}
\end{equation}
Thus, substituting into \eqnref{eq:una} we obtain the precision bound
\begin{equation}
\Delta^2 \omega_N\,T \geq \min_{t} \frac{1+ N \ell(t)}{t N^2},
\label{eq:timeopt}
\end{equation}
which in the case of semigroup dynamics coincides with the
asymptotically tight limit derived in \cite{Knysh2014}.

First, beating the SQL-like scaling necessarily requires $\lim_{N\rightarrow \infty} t_{\t{opt}}(N)\!=\!0$.
Assume on the contrary that the optimal evolution time attains some $t_\infty\!>\!0 $ as $N\!\rightarrow\!\infty$.
Then, inspecting \eqnref{eq:l_def}, one sees that the ``no full-revival'' assumption $\eta_\perp(t_\infty)\!<\!1$,
along with the CPTP constraints,
implies $\ell(t_\infty)\!>\!0$ and hence \eqnref{eq:timeopt} directly restricts the precision to
asymptotically follow $1/N$. As a consequence, we can focus on the short-time regime
and expand $\eta_\perp(t)$, $\eta_\parallel(t)$ as in \eqnref{eq:cov_coeffs}
to get $\ell(t) \!=\! 2 \alpha_\perp t^{\beta_\perp} \!-\! \frac{1}{2} \alpha_\parallel t^{\beta_\parallel} \!+\! o(t^{\beta_\perp})$.
From CPTP constraints it follows that  $\beta_\perp \!\leq\! \beta_\parallel$ and if $\beta_\perp\!=\!\beta_\parallel$
then additionally $\alpha_\parallel \!\leq\! 2 \alpha_\perp$.
Hence, up to the leading order:
\begin{equation}
\label{eq:ellell}
\ell(t)=1 -\alpha_l t^{\beta_\perp}\!+ o(t^{\beta_\perp}),\quad  \alpha_l =
\begin{cases} 2 \alpha_\perp &  \beta_\perp < \beta_\parallel; \\
2 \alpha_\perp - \frac{\alpha_\parallel}{2} & \beta_\perp = \beta_\parallel.
\end{cases}
\end{equation}
Plugging the above expansion into \eqnref{eq:timeopt}, we find that
its minimum is reached for
\begin{equation}\label{eq:ttopt}
t_{\t{opt}}(N) \overset{N\to\infty}{=}(\alpha_l (\beta_\perp - 1) N)^{-1/\beta_\perp},
\end{equation}
which yields the bound \eref{eq:step2} with $\alpha\!=\!\alpha_l$, so that
correctly \eqnsref{eq:fincases2}{eq:ellell} coincide for $\kappa\!=\!0$.

\textit{Attaining the ultimate precision.---}%
As \eqnref{eq:CE_bound} provides us only with an upper limit on the QFI, we still must investigate the
tightness of bound \eref{eq:step2}. Yet, note that also the QCRB \eref{eq:una} itself is guaranteed to
be saturable only in the limit of infinite independent experimental repetitions $T/t\!\to\!\infty$.
This issue is particularly important in the noiseless case,  when the minimisation of the MSE \eref{eq:una}
over $t$ yields $t_{\t{opt}}\!=\!T$, indicating that a single experimental shot consuming all time-resources
should be performed \cite{Escher2011}. The QCRB is then not saturable, what can be demonstrated by
means of rigorous Bayesian approach \cite{Berry2000, *Jarzyna2015}. Fortunately, in the presence of IIC noise
the optimal single-shot duration, $t_\t{opt}$, is independent of $T$ and decays as $N^{-1/\beta_\perp}$ with $N$, see \eqnref{eq:ttopt},
so that $T/t$ always diverges as $N \!\rightarrow\!\infty$. Thus, only due to noise we may assure that
for any $N$ there exists large enough $T$ for which the QCRB is saturable.

We now show that the scaling exponent in \eqnref{eq:step2} is always correct
and it is only the constant $D$ that in some cases may be underestimated.
Consider a GHZ state $\ket{\psi_\t{\tiny GHZ}}\!=\!\frac{1}{\sqrt{2}}(\ket{0}^{\ot N} + \ket{1}^{\ot N})$.
Thanks to its simple structure, the expression for its QFI w.r.t.~the estimated $\omega$ may be analytically derived:
\begin{equation}
F_\t{Q}[\Lambda_\omega^{\ot N}[\psi_\t{\tiny GHZ}^N]]
=
\frac{t^2 N^2 \eta_\perp^{2 N}}{2^{-1-N} \left(A_{-,-}^N+A_{+,-}^N+A_{-,+}^N+A_{+,+}^N\right)},
\end{equation}
with $A_{\pm,\pm}\!=\!1\pm\eta_\parallel\pm\kappa$. Focusing again for simplicity on
unital maps with $\kappa\!=\!0$ (see \appref{app:GHZ} for the general case),
expanding the above formula for short times and using optimal
$t_{\t{\tiny GHZ}}(N) \!=\! 1/(\alpha_l \beta_\perp N)^{1/\beta_\perp}$ that minimises asymptotically
the QCRB \eref{eq:una} for the GHZ-based scenario, we arrive at
\begin{equation}
\lim_{N \rightarrow \infty} \frac{\Delta^2 \omega^\t{\tiny GHZ}_N \,T}{N^{-(2 \beta_\perp-1)/\beta_\perp}}  \;=\; (\alpha_l \beta_\perp \ee)^{1/\beta_\perp}.
\label{eq:GHZ_as_const}
\end{equation}
For the semigroup case ($\beta_\perp\!=\!1$) the asymptotic coefficient \eref{eq:GHZ_as_const} differs 
by a factor $\ee$ from $D$ of \eqnref{eq:step2}---a known fact for the pure dephasing model
\cite{Huelga1997, Escher2011} which may be remedied by replacing GHZ with spin-squeezed states \cite{Ulam2001}---%
yet the discrepancy decreases with increasing $\beta_{\perp}$. Crucially, \eqnref{eq:GHZ_as_const} proves
that the $1/N^{(2 \beta_\perp-1)/\beta_\perp}$  scaling of the MSE predicted by \eqnref{eq:step2} is indeed
always achievable when $\kappa\!=\!0$, however, such claim applies to all phase covariant maps, see \appref{app:GHZ}.

\textit{Role of non-Markovianity and Zeno regime.---}%
We have shown that by going beyond the semigroup regime one can overcome
the SQL for a relevant class of open system dynamics. A natural question is whether
non-Markovian features are of some relevance. Since non-Markovianity is typically associated with
backflow of information to the system of interest \cite{Breuer2009,*Breuer2015,Rivas2010,*Rivas2014},
one may think that such recovered information (also about the estimated parameter) could be 
advantageous for metrological purposes, possibly leading to improved scalings of precision.
Our results clearly indicate that this is \emph{not} the case. As any measurement strategy outside the short-time
regime will be asymptotically bounded by a $1/N$ scaling, to beat the SQL
one \emph{must} perform measurements on shorter and shorter timescales as $N\!\to\!\infty$,
whatever the subsequent memory effects are. The attainable asymptotic precision
is then \emph{fully} dictated by the time-inhomogeneous, i.e., non-semigroup, nature of the dynamics.

The characterisation of non-semigroup dynamics
is a complex task, which calls for a detailed knowledge of the environmental properties,
as well as the interaction mechanism \cite{Breuer2002}. Yet, a general property of any
evolution derived exactly from the global (system+environment) unitary dynamics
is the quadratic decay of the survival probability at short timescales---%
the emergence of the so-called \emph{quantum Zeno regime} \cite{Facchi2008,*Pascazio2014}.
In \appref{app:driving_H}, we explicitly show that for any phase covariant $\Lambda_{\omega}(t)$
such quadratic decay implies that $\beta_\perp\!=\!2$ and \eqnref{eq:step2} then reduces
to $\lim_{N \rightarrow \infty} \Delta \omega^2_N T\,N^{3/2}\!\geq\!\sqrt{\alpha}$.
Thus, we can conclude that the ultimate $1/N^{3/2}$ precision scaling---provably attainable---is 
a general feature of \emph{any} reduced dynamics exhibiting the Zeno regime and phase covariance.
In particular, if we restrict to the specific case of pure dephasing,
we provide further confirmation of the conjecture made in \cite{Chin2012}
and also recently proved in \cite{Macieszczak2015}.

\begin{figure}[!t]
\includegraphics[width=\columnwidth]{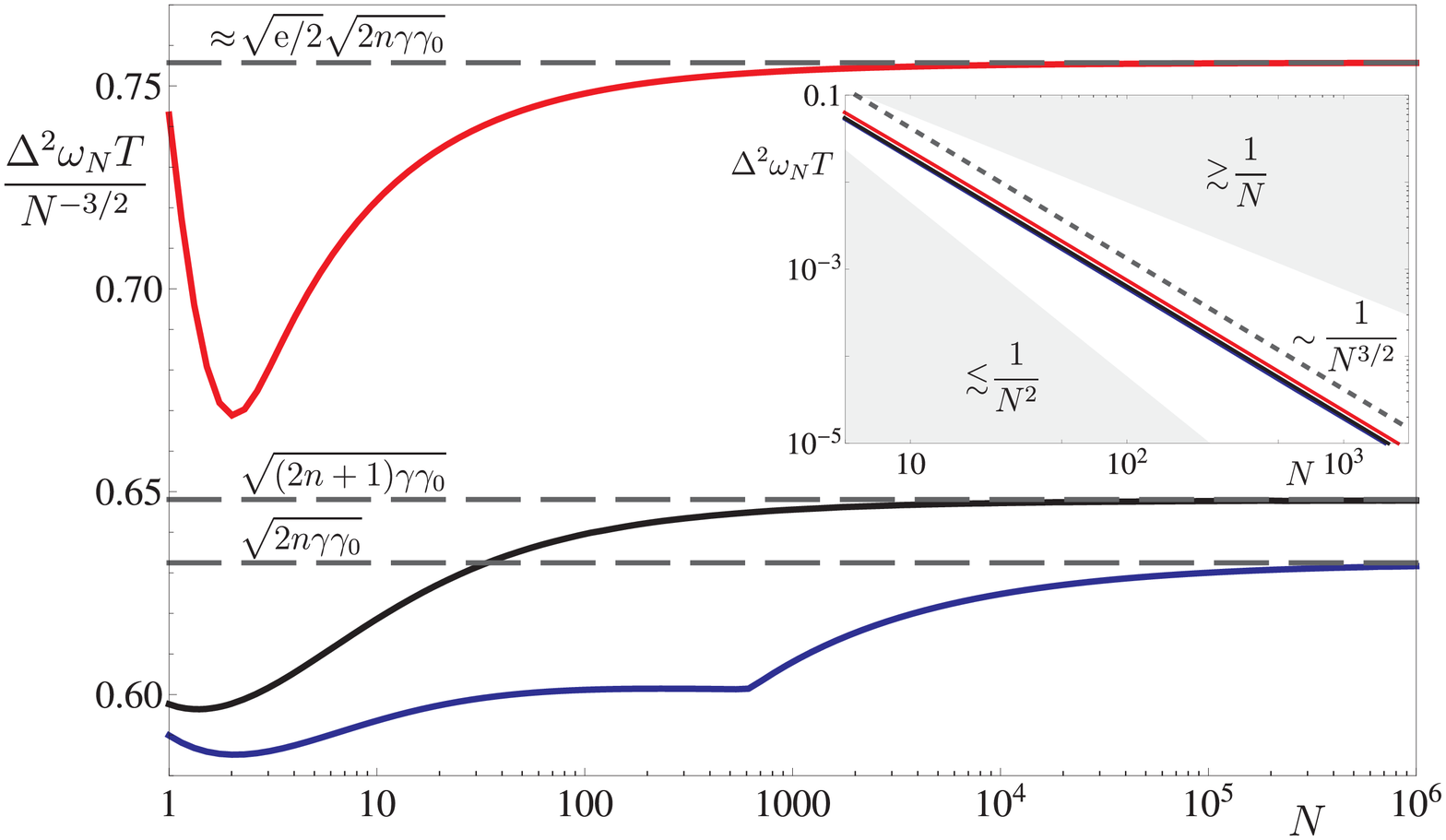}
\caption{\textbf{Attainable precisions for the Shabani-Lidar post-Markovian noise model}
($\gamma\!=\!0.2$, $\gamma_0\!=\!0.1$ and $n\!=\!10$). From bottom to top, we plot
the obtained semi-analytic (\emph{blue}) and fully numerical SDP-based (\emph{black})
precision bounds, in \eqnsref{eq:timeopt}{eq:CE_bound}, respectively, as well as the exact MSE attainable with the GHZ states (\emph{red}).
The inset confirms that all three quickly saturate the $1/N^{3/2}$ scaling with $N$.}
\label{fig:prec_plots}
\end{figure}

\textit{Shabani-Lidar post-Markovian noise model.---}%
To demonstrate the applicability of our methods for general phase covariant dynamics,
we consider the \emph{post-Markovian model of Shabani and Lidar} \cite{Shabani2005} (SL),
that has been widely used to study non-semigroup evolutions and their non-Markovian properties
\cite{Maniscalco2006,*Maniscalco2007,*Mazzola2010}. The SL master equation \eref{eq:step1me}
contains all the emission, excitation and dephasing contributions,
yet it is fully described by three parameters:~$\gamma_0$--dissipation constant,
$\gamma$--the effective memory rate, and $n$--the mean number of excitations in the reservoir.
In \appref{app:SLmodel}, we show the short-time expansions \eref{eq:cov_coeffs} of
its noise parameters ($\eta_\perp$, $\eta_\parallel$ and $\kappa$) to be quadratic in $t$
($\beta_\perp\!=\!\beta_\para\!=\!\beta_\kappa\!=\!2$) with coefficients:~$\alpha_\para\!=\!2\alpha_\perp\!=\!(2n+1)\gamma_0\gamma/2$
and $\alpha_\kappa\!=\!-\gamma \gamma_0/2$. The SL model therefore exhibits the natural Zeno regime, which
imposes the $1/N^{3/2}$ asymptotic precision scaling with $D\!=\!\sqrt{2n\gamma \gamma_0}$ in
\eqnref{eq:step2}. On the other hand, given that the model yields a non-unital qubit map, the GHZ-achievable
asymptotic constant \eqnref{eq:GHZ_as_const} is computed numerically and can
be approximated by $\sqrt{\ee/2}\,D$ (see \appref{app:SLmodel}). The SL model is 
well suited to verify the performance of our methods at finite $N$. In \appref{app:SLmodel}, we derive the corresponding analytic
expressions for the bound \eref{eq:timeopt} and the GHZ-attainable precision, both of which we numerically optimise 
over $t$ for each $N$. We plot the results in \figref{fig:prec_plots} to show that they converge to the correct 
asymptotic, analytical expressions. Furthermore, we compare the obtained semi-analytical bound \eref{eq:timeopt} with its
fully numerically optimised SDP-based version \eref{eq:CE_bound}, which we find to asymptotically achieve only a slightly
tighter $\sqrt{(2n\!+\!1)\gamma \gamma_0}$ constant. The inset of \figref{fig:prec_plots} clearly confirms
the attainability of the $1/N^{3/2}$ precision scaling.

\textit{Conclusions.---}%
We have derived a novel limit on the attainable precision in frequency
estimation, which holds for all forms of phase covariant uncorrelated noise.
Our results show that, despite the noiseless HL not being within reach, by exploiting 
the non-semigroup, time-inhomogeneous system dynamics arising at short-times in 
the Zeno regime, the asymptotic SQL-like scaling of precision can be beaten.
Any measurement strategies performed on longer timescales are
always ultimately limited by the SQL, irrespectively of the possible non-Markovian
effects exhibited by the evolution. We leave it as an open question whether the asymptotic
precision can be further improved by means of general active-ancilla assisted schemes
\cite{Demkowicz2014}, where the interplay between the multi-step unitary operations
and the memory effects in the probes evolution has to be carefully treated.

\begin{acknowledgments}
We acknowledge enlightening discussions with Bogna Bylicka.
This work has been supported by Spanish Ministry National Plan FOQUS,
Generalitat de Catalunya (SGR875), EU MSCA Individual Fellowship Q-METAPP,
and EU FP7 IP project SIQS co-financed by the Polish
Ministry of Science and Higher Education. This work has also received funding from the
European Union's Horizon 2020 research and innovation programme under the QUCHIP project GA no. 641039
\end{acknowledgments}

\bibliographystyle{apsrev4-1}
\bibliography{zeno_metro_ar}

\appendix

\section{Dynamical maps -- matrix and time-local master equation representations}
\label{app:cptp_maps_reps}

In this section, we briefly recall the essential features of dynamical maps we use
throughout our work and fix the notation. In particular, we focus on the
matrix representation of the completely positive trace preserving (CPTP) linear maps,
which is usually exploited to describe the dynamics of open quantum systems%
---see, e.g., \cite{Bengtsson2006, Ruskai2002,Andersson2007,Asorey2009,Chruscinski2010,Smirne2010}.

\subsection{Matrix representation of completely positive maps}

Given a finite dimensional Hilbert space, $\mathcal{H}\!=\!\mathbbm{C}^d$, the set
$\mathcal{L}(\mathbbm{C}^d)$ of linear operators on $\mathbbm{C}^d$ also forms a Hilbert space
equipped with a Hilbert-Schmidt scalar product:
\begin{equation}
\langle \sigma, \tau  \rangle = \t{Tr}\left\{\sigma^{\dag} \tau\right\},
\end{equation}
where $\sigma, \tau \in \mathcal{L}(\mathbbm{C}^d)$. Moreover, given the set
$\mathcal{L}\mathcal{L}(\mathbbm{C}^d)$ of linear maps acting on $\mathcal{L}(\mathbbm{C}^d)$,
the Hilbert-Schmidt scalar product naturally induces
a one-to-one correspondence between $\mathcal{L}\mathcal{L}(\mathbbm{C}^d)$
and the set of $d^2\!\times\!d^2$ matrices. Explicitly,
for any basis $\left\{\varsigma_{j}\right\}_{j = 1, \ldots d^2}$ in $\mathcal{L}(\mathbbm{C}^d)$
that is orthonormal with respect to such scalar product, i.e.,
$\langle \varsigma_{j}, \varsigma_{l} \rangle = \delta_{j l}$, one has
\begin{equation}
\Lambda [\tau] = \sum^{d^2}_{j l = 1} {\sf\Lambda}_{j l} \langle \varsigma_{l} , \tau \rangle \varsigma_{j},
\qquad {\sf \Lambda}_{j l} = \langle \varsigma_{j}, \Lambda[\varsigma_{l}]\rangle,
\label{eq:mr}
\end{equation}
for any $\tau\!\in\!\mathcal{L}(\mathbbm{C}^d)$
and any linear map $\Lambda\!\in\!\mathcal{L}\mathcal{L}(\mathbbm{C}^d)$. In this way,
any map $\Lambda$ is univocally associated with a matrix ${\sf{\Lambda}}$ with elements ${\sf \Lambda}_{j l}$.
It is easy to see that the composition of two maps, $\Lambda \circ \Phi$, corresponds to
the matrix product ${\sf {\Lambda}} {\sf{\Phi}}$, and thus the inverse of the map $\Lambda$ is represented
by the inverse matrix ${\sf \Lambda}^{-1}$.
Considering a basis $\left\{\varsigma_{j} \right\}_{j = 0, \ldots d^2-1}$
such that $\varsigma_0\!=\!\mathbbm{1}/ \sqrt{d}$
and, for all $j\!\geq\!1$, $\varsigma_{j}$ are orthonormal traceless self-adjoint operators,
the map $\Lambda$ is trace-preserving \emph{if and only if} its matrix representation can be written as
\begin{equation}
{\sf{ \Lambda}} = \left(\begin{array}{cc}
    1 & {\bf 0}\\
    {\bf m } & {\sf M}
    \end{array}
     \right),
\label{eq:mrtp}
\end{equation}
where ${\bf 0}$ is a row vector made of 0s, ${\bf m }$ is a column vector,
and ${\sf M}$ is a $(d^{2}\!-\!1)\!\times\!(d^{2}\!-\!1)$ matrix. Furthermore, $\Lambda$ is
hermiticity-preserving (i.e., maps hermitian operators onto hermitian operators) \emph{if and only if}
${\bf m }$ and ${\sf M}$ are real.

In the case of qubit ($\mathcal{H} \!=\! \mathbbm{C}^2$) maps,
their matrix form has a simple geometrical interpretation, relying
on the Bloch-ball representation of the qubit states.
Given the basis $\left\{\mathbbm{1}/\sqrt{2}, \sigma_{j}/\sqrt{2}\right\}_{j = x, y, z}$
on $\mathcal{L}(\mathbbm{C}^2)$, where the $\sigma_j$s are the usual Puali matrices,
any statistical operator $\varrho$ on $\mathbbm{C}^2$ can be decomposed as
\begin{equation}
\varrho = \frac{1}{2}\left(\mathbbm{1} + \bf{v} \cdot \boldsymbol \sigma \right), \label{eq:blsp}
\end{equation}
where ${\bf v}$ is the 3-dimensional real vector with elements
 $v_j \!=\! \t{Tr}\!\left\{{\sigma_j} \varrho\right\}$ and such that $|{\bf v}| \leq 1$, while $\boldsymbol \sigma$
is a vector of Pauli matrices. Such decomposition defines the well-known one-to-one correspondence
between the set of the statistical operators on $\mathbbm{C}^2$ and the 3 dimensional closed real ball of radius
one centered at the origin, i.e., the Bloch ball. Moreover, any linear map with matrix representation as in \eqnref{eq:mrtp}
just yields
\begin{equation}\label{eq:bbll}
\Lambda[\varrho] = \frac{1}{2}\left(\mathbbm{1} +  (\bf{m} + {\sf M} \bf{v}) \cdot \boldsymbol \sigma \right).
\end{equation}
In particular, the action of the map corresponds to an affine transformation of the Bloch ball,
${\bf v} \rightarrow {\bf m} + {\sf M} \bf{v} $, where ${\bf m }$ describes the translations, while
${\sf M}$ describes:~ rotations, contractions and reflections about the three orthogonal axis
(as can be seen after performing the singular value decomposition \cite{Ruskai2002}).
In addition, the matrix representation allows for a clear geometrical characterisation of the complete
positivity of qubit maps \cite{Ruskai2002}. This is done by studying the positivity
of the Choi matrix $\Omega_\Lambda$ associated with
the map $\Lambda$ \cite{Bengtsson2006}. For any orthonormal basis $\left\{\ket{u}_j\right\}_{j=1, \ldots d}$ in $\mathbbm{C}^d$
one can define the Choi matrix as
\begin{equation}\label{eq:choi}
\Omega_\Lambda  = \left(\begin{array}{cccc}
    \Lambda(e_{11}) & \Lambda(e_{12}) & \ldots &  \Lambda(e_{1d}) \\
   \Lambda(e_{21})  & \Lambda(e_{22})  & \ldots & \Lambda(e_{2d}) \\
   \vdots & \vdots &  & \vdots\\
   \Lambda(e_{d1})  & \Lambda(e_{d2})  & \ldots & \Lambda(e_{dd}),
    \end{array}
     \right),
\end{equation}
with $e_{jk} \!=\! \ket{u_j} \bra{u_k} $, so that the CP of $\Lambda$
is equivalent to the positivity of the matrix $\Omega_\Lambda$.

\subsection{Time-local master equations of open quantum systems}

Any dynamics of an open quantum systems may be generally described by
a one-parameter family of CPTP maps
$\left\{\Lambda(t)\right\}_{t\geq0}$, where the instance
$t\!=\!0$ just corresponds to the initial time of the evolution.
Thus the initial map must read
\begin{equation}
\label{eq:ci}
\Lambda(0) = \mathbbm{1},
\end{equation}
while the system state at any later times is described by
\begin{equation}
\label{eq:masi}
\varrho(t) = \Lambda(t)[\varrho(0)].
\end{equation}
As discussed, $\Lambda(t)$ can then be specified for any fixed time $t$ with
help of its matrix representation ${\sf\Lambda}(t)$ in \eqnref{eq:mrtp}. 
Yet, we also consider the time-local master equation (TLME) satisfied by 
the state $\varrho(t)$ at any $t$:
\begin{equation}\label{eq:xi}
\frac{\dd}{\dd t} \varrho(t) = \Xi (t) [\varrho(t)].
\end{equation}
Given a one-parameter family $\left\{\Lambda(t)\right\}_{t\geq0}$ defining the dynamics,
its corresponding TLME can be formally defined as
\cite{Andersson2007,Chruscinski2010,Smirne2010}:
\begin{equation}\label{eq:xxi}
\Xi (t) = \frac{\dd \Lambda(t)}{\dd t} \circ \Lambda(t)^{-1}.
\end{equation}
It is then clear that the matrix representation of the dynamical maps defined by \eqnref{eq:mr} can
be further exploited to get the matrix associated with the time-local generator $\Xi(t)$ reading:
\begin{equation}
{\sf \Xi} (t) = \frac{\dd{{\sf \Lambda}(t)}}{\dd t} {\sf \Lambda}(t)^{-1}.
\end{equation}
Applying \eqnref{eq:mr} to $\Xi(t)$, one may thus get an explicit form of the TLME, which
for any trace- and hermiticity-preserving dynamics can be written as \cite{Gorini1976}:
\begin{eqnarray}
\label{eq:mett}
\frac{\dd\varrho(t)}{\dd t} &=& \Xi (t) [\varrho(t)]= - \ii \left[ H(t), \varrho(t) \right] + \\
&&\!\!\!\!\!\!\!\!\!\!\!
+\sum^{d^2-1}_{j = 1}\! \gamma_{j}(t)\!\left(\!L_j(t) \varrho(t) L_j^{\dag}(t) -
\frac{1}{2}\! \left\{ L^{\dag}_{j}(t) L_j(t), \varrho(t) \right\}\!\right)\!, \nonumber
\end{eqnarray}
where $H(t)=H(t)^{\dag}$ is the Hamiltonian contribution and the $L_j(t)$ are the (linear independent) Lindblad operators.
The rates $\gamma_j(t)$ can be in general time-dependent and, importantly, one may deal with 
a well-defined CPTP evolution also in the presence of rates taking on negative values.
In the next paragraphs, we will explicitly discuss the connection between the master equation
and non-Markovianity in quantum dynamics.
Customarily, one assumes that the time derivative of the dynamical map considered above
always exists and is continuous, or equivalently that all the matrix elements of ${\sf \Lambda}(t)$
are $C^1(\mathbbm{R}_0^+)$ functions. However, let us mention that there exist interesting
dynamics exhibiting time instants at which the inverse of
the dynamical map $\Lambda(t)^{-1}$ cannot be defined, typically due to some of the rates
$\gamma_{j}(t)$ being divergent
\footnote{Owing to adequate constraints, the TLME may still be defined at such instances despite
$\Lambda(t)$ being then non-invertible \cite{Andersson2007}.}. Our analysis will also cover these dynamics.

Finally, let us clarify that when considering $N$-qubit systems ($\mathcal{H}\!=\!\mathbbm{C}^{d}$
with $d\!=\!2^N$), we denote by $O^{N}$ a general linear operator in $\mathcal{L}(\mathbbm{C}^{2^N})$,
while by $O^{\ot N}$ an $N$-fold tensor product of a given single-qubit operator $O$, i.e.,
$O^{\ot N}\!\equiv\!O\ot \ldots \ot O$ ($N$ times). An analogous notation is used when describing linear
maps from $\mathcal{L}\mathcal{L}(\mathbbm{C}^{2^N})$. In particular, given a unitary operator
$U^{N}$ in $\mathcal{L}(\mathbbm{C}^{2^N})$, we denote by $\mathcal{U}^{N}$ the unitary map
(operator in $\mathcal{L}\mathcal{L}(\mathbbm{C}^{2^N})$) defined as
\begin{equation}
\mathcal{U}^{N}[\rho] = U^{N} \rho (U^{N})^{\dag}.
\end{equation}
Lastly, let us note that given two linear maps $\Xi$ and $\Lambda$ we denote their
commutator and anti-commutator by $\left[\Xi, \Lambda\right] = \Xi \circ \Lambda -  \Lambda \circ \Xi$
and $\left\{\Xi, \Lambda\right\} = \Xi \circ \Lambda +  \Lambda \circ \Xi$
respectively.

\subsection{Quantum Markovianity}
Here, we briefly recall the distinction between Markovian
and non-Markovian dynamics for open quantum systems, in particular w.r.t.~the TLME in \eqnref{eq:mett}
and the notion of time-homogeneity;
for a more detailed treatment the reader is referred to the recent reviews
in \cite{Breuer2009,Rivas2010}.

From a physical point of view,
the dynamics of an open quantum system is Markovian if
the memory effects due to its interaction with the environment can be neglected,
typically due to a definite separation in the evolution time-scales of, respectively, the open system and the environment  \cite{Breuer2002}.
More precisely, quantum Markovianity can be formulated in terms of the divisibility
properties of the dynamical maps $\Lambda(t)$.
One defines the dynamics to be \emph{divisible} if
\begin{equation}\label{eq:div}
\Lambda(t) = \Lambda(t,s) \Lambda(s),
\end{equation}
for any $t \geq s \geq 0$.
Note that any dynamics such that $\Lambda(t)^{-1}$ exists at every time $t$
is divisible, as can be seen by simply setting $\Lambda(t,s) = \Lambda(t) \circ \Lambda(s)^{-1}$,
but $\Lambda(t,s)$ is \emph{not} in general a CP map.
The linear maps $\Lambda(t,s)$
are usually referred to as the propagators of the dynamics and they can be expressed in terms
of the TLME in \eqnref{eq:mett} as \cite{Rivas2012}:
\begin{equation}
\Lambda(t,s) = T_{\leftarrow} \exp\left(\int_s^t \!\dd s\; \Xi(s)\right),
\end{equation}
where one has the identification $\Lambda(t,0) = \Lambda(t)$.
If the propagators depend only on the difference
between their time arguments, i.e., $\Lambda(t,s) = \Lambda(t-s,0)$ for any $t \geq s \geq 0$,
the dynamics is said to be \emph{time-homogeneous}.
One can easily see that this precisely corresponds to the case in which the TLME
has constant coefficients. Furthermore, in this case, the dynamical
maps satisfy
\begin{equation}
\Lambda(t) = \Lambda(t-s) \Lambda(s),
\end{equation}
for any $t \geq s \geq 0$, which is the well-known \emph{semigroup} composition law.
The most general form of the (bounded) generator of a semigroup of CPTP maps was characterized
by Gorini, Kossakowski, Sudarshan and Lindblad in \cite{Gorini1976},
and is given, in the finite dimensional case, by \eqnref{eq:mett}
with constant positive coefficients $\gamma_j \geq 0$.

Quantum semigroup dynamics have been identified as the time-homogeneous Markovian 
dynamics in the quantum setting, both because of the analogy with the semigroup composition 
law for the transition probabilities of classical time-homogeneous Markovian stochastic processes,
and because they describe satisfactorily the dynamics of open quantum systems when one can 
fully neglect the memory effects encoded into the environmental multi-time correlation functions \cite{Breuer2002}.
A natural way to extend the definition of quantum Markovianity also to \emph{time-inhomogeneous}
dynamics is then to say that a given dynamics is Markovian when it is \emph{CP-divisible} \cite{Rivas2010},
i.e., not only \eqnref{eq:div} holds, but also the maps $\Lambda(t,s)$ are CPTP for any $t \geq s \geq 0$.
Furthermore, the dynamics generated by $\Xi(t)$ in \eqnref{eq:mett}
is CP-divisible if and only if all the rates are non-negative functions of time, i.e., $\gamma_j(t) \geq 0$
for any $j$ and $t$.

As a last remark, let us stress that different and non-equivalent definitions of Markovianity have been introduced.
Nevertheless, the conclusions of this work do not depend on the definition exploited, as they only rely 
on the distinction between time-homogenous and time-inhomogeneous dynamics.

\section{Phase covariant dynamics}
\label{app:cov_maps_reps}

In this section, we show explicitly how to characterize
the class of reduced dynamics due to identical independent and phase covariant (IIC) noise.
First we derive the general form of the phase covariant map as given in \eqnref{eq:phasecov_map}
of the main text and then provide its corresponding TLME.

\subsection{Phase covariant maps}

The most general form of IIC maps could be obtained via the
theory of covariant quantum channels \cite{Holevo1993,*Holevo1996},
which classifies the maps commuting with some group representation
via the Choi-Jamiolkowski isomorphism \cite{Bengtsson2006}. However, for the simple case of
$U(1)$-covariant qubit channels, i.e., the \emph{phase covariant qubit channels},
we can directly exploit the simpler tools provided by the matrix representation of dynamical maps
that has been introduced in the previous section.

Let $\Gamma$ be a trace- and hermiticity- preserving linear map in $\mathcal{L}\mathcal{L}(\mathbbm{C}^{2})$
and $\mathcal{U}_{\omega}$ be a unitary map also in $\mathcal{L}\mathcal{L}(\mathbbm{C}^{2})$
fixed by $U_{\omega} = \ee^{-\ii\frac{\omega t}{2} \sigma_z}$ in $\mathcal{L}(\mathbbm{C}^{2})$,
such that
\begin{equation}
\label{eq:comm1}
\left[ \mathcal{U}_{\omega} , \Gamma\right] = 0.
\end{equation}
By \eqnref{eq:mr}, the matrix representation of $\mathcal{U}_{\omega}$
is given by
\begin{equation}
\label{eq:uu}
{\sf{U_{\omega}}} = \left(\begin{array}{cccc}
    1 & 0 & 0 & 0\\
   0 & \cos \omega t &-\sin \omega t & 0\\
   0 & \sin \omega t & \cos \omega t & 0\\
   0 & 0 & 0 & 1
    \end{array}
     \right).
\end{equation}
As clear from the previous section, commutation relation $\left[ \mathcal{U}_{\omega} , \Gamma\right] = 0$ is equivalent to the same relation
between the corresponding matrices $\left[{\sf U}_{\omega},  {\sf \Gamma} \right] = 0$.
Taking the matrix representation ${\sf \Gamma}$ given by \eqnref{eq:mrtp}, with real ${\bf m }$ and ${\sf M}$, the
vanishing commutator with the unitary-map matrix \eref{eq:uu} requires
\begin{eqnarray}
({\sf M}_{12}+{\sf M}_{21}) \sin \omega t = ({\sf M}_{22}-{\sf M}_{11}) \sin \omega t  &=& 0\nonumber\\
{\sf M}_{13}(\cos \omega t-1)-{\sf M}_{23}\sin \omega t &=& 0 \nonumber\\
 {\sf M}_{23}(\cos \omega t-1)+{\sf M}_{13} \sin \omega t &=& 0 \nonumber\\
{\sf M}_{31}(1-\cos \omega t)-{\sf M}_{32}\sin \omega t &=& 0 \nonumber\\
{\sf M}_{32}(1-\cos \omega t)+{\sf M}_{31} \sin \omega t  &=& 0 \nonumber\\
m_{1}(\cos \omega t-1)- m_{2}\sin \omega t &=& 0\nonumber\\
 m_{2}(\cos \omega t-1)+m_{1} \sin \omega t &=& 0 \nonumber,
\end{eqnarray}
so that ${\sf M}_{12} \!=\! -{\sf M}_{21}$, ${\sf M}_{22} \!=\! {\sf M}_{11}$, while
${\sf M}_{13} \!=\! {\sf M}_{23} \!=\! {\sf M}_{31} \!=\! {\sf M}_{32} \!=\!m_{1}\!=\!m_2\!=\!0$.
Thus, the map $\Gamma$ is restricted to have the form of \eqnref{eq:mrtp} with just non-zero $m_3$
yielding a translation along the $z$ axis and the matrix ${\sf M}$ describing a rotation around the $z$ axis
augmented by a contraction of the Bloch ball, such that the resulting ellipsoid has equal axes along $x$ and
$y$ with potential reflection about the $xy$ plane---see \figref{fig:cov_maps} of the manuscript and \eqnref{eq:bbll}.
Using the singular value decomposition \cite{King2001} and identifying the translation $m_{3}$ with
$\kappa$, one obtains:
 \begin{equation}
\label{eq:step1}
\sf{\Gamma}= \mat{cccc}{
1 & 0 & 0 & 0\\
0 & \eta_{\perp} \cos\theta &  -\eta_{\perp} \sin\theta & 0 \\
0 & \eta_{\perp} \sin\theta& \eta_{\perp} \cos\theta & 0 \\
\kappa & 0 & 0 & \eta_{\parallel}
}
\end{equation}
with $\eta_{\perp}, \eta_{\parallel}, \kappa, \theta$ as defined in the main text
and pictorially described in \figref{fig:cov_maps} therein. Finally, by multiplying
${\sf \Gamma}$ and ${\sf U}_\omega$ (that commute with one another) we obtain the matrix form
of $\Lambda_{\omega}$ which is valid for each element of the $t$-parametrised family of maps 
describing the dynamics stated in \eqnref{eq:phasecov_map} of the main text:
\begin{equation}
\label{eq:phasecov_map2}
{\sf\Lambda}_{\omega}= \mat{cccc}{
1 & 0 & 0 & 0\\
0 & \eta_{\perp} \cos\phi&  -\eta_{\perp} \sin\phi & 0 \\
0 & \eta_{\perp}\sin\phi & \eta_{\perp} \cos\phi & 0 \\
\kappa & 0 & 0 & \eta_{\parallel},
}
\end{equation}
with $\phi=\omega t+\theta$.

For future convenience, let us note that the Choi matrix associated with the map described
by \eqnref{eq:phasecov_map2} is given in the canonical basis (see \eqnref{eq:choi}) by
\begin{equation}
\Omega_{\Lambda_\omega} =  \mat{cccc}{
\tfrac{1+\eta_\parallel+\kappa}{2} & 0 & 0 & \eta_\perp \ee^{-\ii \phi} \\
0 & \tfrac{1-\eta_\parallel+\kappa}{2} & 0 & 0 \\
0 & 0 & \tfrac{1-\eta_\parallel-\kappa}{2} & 0 \\
\eta_\perp \ee^{\ii \phi} & 0 & 0 & \tfrac{1+\eta_\parallel - \kappa}{2}},
\label{eq:choii}
\end{equation}
so that $\Lambda_{\omega}$ is CP if and only if:
\begin{align}
&\eta_\parallel \pm \kappa \leq 1, \nonumber\\
&1+ \eta_\parallel \geq \sqrt{ 4 \eta_\perp^2 + \kappa^2}.\label{eq:cpc}
\end{align}
The above equations clearly imply that $-1\!\leq\!\kappa\!\leq\!1$,
$-1\!\leq\!\eta_\parallel\!\leq\!1$ and without loss of generality we additionally
restrict  $0\!\leq\!\eta_\perp\!\leq\!1$ (negative $\eta_\perp$ simply correspond
to an additional $\pi$ rotation around the $z$ axis).
For this class of dynamical maps, the unitary part does not affect the CP constraints \eref{eq:cpc},
yet they will play an important role in the following analysis of \appref{app:Zeno_bound}.
Finally, by considering the eigenvectors of the Choi matrix \eref{eq:choii},
$\Omega_{\Lambda_\omega}\!=\!\sum_i\ket{K_i}\!\bra{K_i}$
with $\ket{K_i}\!=\!(K_i\ot\openone)\sum_{j=0}^1\ket{j,j}$, one may define
the canonical Kraus operators that satisfy $\Lambda_{\omega}[\varrho]=\!\sum_i  K_i \varrho K_i^{\dag}$
\cite{Bengtsson2006}:
\begin{align}\label{eq:Kraus_ops}
K_1 &= \sqrt{\tfrac{1-\eta_\parallel+\kappa}{2}} \mat{cc}{0 & 1 \\ 0 & 0} \nonumber\\
K_2 & = \sqrt{\tfrac{1-\eta_\parallel-\kappa}{2}} \mat{cc}{0 & 0 \\ 1 & 0} \nonumber\\
K_3 & = \sqrt{\lambda_+} \mat{cc}{\cos\vartheta & 0 \\ 0 & \sin\vartheta \,\ee^{\ii \phi}} \nonumber\\
K_4 & = \sqrt{\lambda_-} \mat{cc}{-\sin\vartheta & 0 \\ 0 & \cos\vartheta \,\ee^{\ii \phi}},
\end{align}
where
\begin{align}
\lambda_{\pm} & = \frac{1 + \eta_\parallel \pm \sqrt{\kappa^2 + 4\eta_\perp^2}}{2} \nonumber\\
\cot \vartheta & = \frac{\kappa + \sqrt{\kappa^2 + 4 \eta_\perp^2}}{2 \eta_\perp}. \label{eq:cot(theta)}
\end{align}

In particular, note that typical qubit channels (see \cite{Bengtsson2006}) correspond
to special instances of the map $\Lambda_{\omega}$, i.e.:~\emph{pure dephasing}
is recovered by setting $\eta_\para\!=\!1$, $\kappa\!=\!0$ and considering
$\eta_\perp>\!0$; for \emph{isotropic depolarisation} (white local noise)
$\eta_\para\!=\!\eta_\perp\!>\!0$, $\kappa=0$; whereas
\emph{amplitude damping} corresponds to $0\!\leq\!\kappa\!\le\!1$, $\eta_\para\!=\!1-\kappa$
and $\eta_\perp\!=\!\sqrt{1-\kappa}$.

The \emph{unital} phase covariant transformations, i.e.,
ones that preserve identity, $\Lambda[\tfrac{1}{2}\openone]\!=\!\tfrac{1}{2}\openone$,
consist of all maps with no displacement:~$\kappa\!=\!0$.
Crucially, in the unital case the Kraus operators
\eref{eq:Kraus_ops}  significantly simplify
with $\vartheta\!=\!\pi/4$. In fact, this will allow us to construct in \appref{app:CEbound_cov}
the quantum-metrology precision bounds in an analytic form, directly basing on the finite-$N$ CE
methods introduced in \cite{Kolodynski2013,*Kolodynski2014}. For the non-unital case,
in order to retain analyticity of the results, we use the convexity property of such bounds
w.r.t.~mixing of dynamical maps, which we explicitly prove in \appref{app:CEbound_conv}.

In summary, given a \emph{phase covariant dynamics}---%
a one-parameter family of CPTP maps $\left\{\Lambda_{\omega}(t)\right\}_{t\geq0}$
for which at every time $t$ the map $\Lambda_{\omega}(t)$ can be decomposed
into a unitary $\omega$-encoding $\mathcal{U}_{\omega}$ and an $\omega$-independent noise 
term $\Gamma(t)$, such that the two commute---%
we arrive at the general form of the evolution given by \eqnref{eq:phasecov_map2} for any time $t$, i.e.,
\eqnref{eq:phasecov_map} stated in the main text. As said, we assume throughout the work the matrix elements of dynamical maps considered
to be smooth in $t$, what is assured by $\eta_{\parallel}(t), \eta_{\perp}(t), \kappa(t)$ and $\theta(t)$
being real functions of class $C^1(\mathbbm{R}_0^+)$. Furthermore, the initial conditions
$\kappa(0)\!=\! \theta(0) \!=\! 0, \eta_{\perp}(0)\!=\!\eta_{\parallel}(0)\!=\!1$ guarantee
that \eqnref{eq:ci} holds, while conditions of \eqnref{eq:cpc} ensure
that any map considered is indeed CP.

\subsection{Time-local master equation of a phase covariant dynamics}
The IIC dynamics can be equivalently characterized by investigating the TLME
associated with the state $\varrho(t)$ in \eqnref{eq:masi},
as we explicitly show below.

Applying \eqnref{eq:xxi}  to the family of maps  $\left\{\Lambda_{\omega} (t)\right\}_{t \geq0}$
specified by \eqnref{eq:phasecov_map2} and using \eqnref{eq:mr} to obtain an explicit form of
the TLME \cite{Smirne2010}, one ends up with \eqnref{eq:step1me} stated in the main text,
i.e., \eqnref{eq:mett} with the time-local ($\omega$-dependent) generator:
\begin{eqnarray}
\Xi_{\omega}(t) &\equiv&  -\frac{i}{2} (\omega + h(t)) [\sigma_z,  \varrho(t)] \nonumber\\
&&+ \gamma_+(t)\left(\sigma_+ \varrho(t) \sigma_- - \frac{1}{2} \left\{\sigma_- \sigma_+, \varrho(t)\right\}\right)  \nonumber\\
&& + \gamma_-(t)\left(\sigma_- \varrho(t) \sigma_+ - \frac{1}{2} \left\{\sigma_+ \sigma_-, \varrho_{\omega}(t)\right\}\right) \nonumber\\
&&+ \gamma_z(t) \left(\sigma_z \varrho(t) \sigma_z -  \varrho(t) \right).
\label{eq:tlg}
\end{eqnarray}
In particular, one finds
\begin{eqnarray}
h(t) &=&\theta'(t)\nonumber\\
\gamma_+(t) &=& \frac{1}{2}\left(\kappa'(t)-\frac{\eta_{\parallel}'(t)}{\eta_{\parallel}(t)}(\kappa(t)+1) \right)   \nonumber\\
\gamma_-(t) &=& -\frac{1}{2}\left(\kappa'(t)+\frac{\eta_{\parallel}'(t)}{\eta_{\parallel}(t)}(1-\kappa(t)) \right)    \nonumber \\
\gamma_z(t) &=&  \frac{1}{4}\left(\frac{\eta_{\parallel}'(t)}{\eta_{\parallel}(t)} - 2 \frac{\eta_{\perp}'(t)}{\eta_{\perp}(t)} \right),
\label{eq:see}
\end{eqnarray}
where we denote the derivative w.r.t.~$t$ by~$f'(t)\!\equiv\!\dd f(t)/ \dd t$.

Let us remark that \eqnref{eq:see} unambiguously defines the TLME except when $\eta_{\perp}(t)\!=\!0$ or $\eta_{\parallel}(t)\!=\!0$,
for which $\Lambda_{\omega}(t)^{-1}$ in \eqnref{eq:xxi} does not exist. As both these parameters are guaranteed to be
equal to identity at $t\!=\!0$, ${\sf \Lambda}_{\omega}(t)$ can become singular only after finite duration of time.
Thus, we may always assure that there exists a short enough range of timescales in which the TLME
obeying \eqnref{eq:see} can be defined.

Crucially, reversing the argument, any TLME defined by the generator \eref{eq:tlg} with real, continuous and bounded
functions $h(t), \gamma_+(t), \gamma_-(t), \gamma_z(t)$, along with the initial condition 
$\Lambda_{\omega}(0)\!=\!\mathbbm{1}$, is uniquely solved by the one-parameter family of 
trace and hermiticity preserving linear maps $\left\{\Lambda_{\omega} (t)\right\}_{t \geq0}$ 
fixed by \eqnref{eq:phasecov_map2}.
Since the coefficients of the master equations are assumed to be bounded, the generator \eref{eq:tlg}
is bounded in norm for any $t$, so that the
solution of the master equation is unique and provided by the Dyson series \cite{Rivas2012}:
\begin{equation}
\Lambda_{\omega}(t) = T_{\leftarrow} \exp\left(\int_0^t \!\dd s\; \Xi_{\omega}(s)\right).
\end{equation}

Consequently, the matrix elements of ${\sf \Lambda}_{\omega}(t)$ are given by the
unique solution to the system of differential equations specified in \eqnref{eq:see}
with initial conditions $\kappa(0)\!=\! \theta(0) \!=\! 0, \eta_{\perp}(0)\!=\!\eta_{\parallel}(0)\!=\!1$.
Hence, most generally:
\begin{eqnarray} \label{eq:soll}
\theta(t) &=&  \int^t_0 \dd s h(s) \\
\kappa(t) &=& \int^t_{0} \dd s  (\gamma_+(s)-\gamma_-(s))\times  \nonumber\\
&&\;\;\times \exp\left(- \int^t_{s} \dd s' (\gamma_+(s')+\gamma_-(s'))\right) \nonumber\\
\eta_{\parallel}(t) &=&   \exp\left(- \int^t_0 \dd s(\gamma_+(s)+\gamma_-(s)) \right) \nonumber\\
\eta_{\perp}(t) &=& \exp\left(-\frac{1}{2}\int^t_0 \dd s(\gamma_+(s)+\gamma_-(s)+4 \gamma_z(s)) \right). \nonumber
\end{eqnarray}

If some of the coefficients in the TLME diverge at instants $\tilde{t}_1\!<\!\tilde{t}_2\!<\!\dots\!<\!\tilde{t}_M$,
one can still find a (unique) solution for all the proper time intervals.
In particular, for any closed interval $[0,\bar{t}_1]$ with $\bar{t}_1\!<\!\tilde{t}_1$
the solution of the master equation can still be written as in \eqnref{eq:soll}.
What is more, it may well happen that the master equation with singularities at
above instants is solved by a (smooth) family of maps well-defined
for every $t$, see for example \cite{Chruscinski2010}.

As a final remark to this section, let us stress the generality of the class of dynamics considered here.
In particular, it originates
from the commutation condition in \eqnref{eq:comm1}, which concerns the action of linear maps in
$\mathcal{L}\mathcal{L}(\mathbbm{C}^{2^N})$ and \emph{not} the operators supported by $\mathcal{L}(\mathbbm{C}^{2^N})$.
In fact, \eqnref{eq:comm1} holds even if the Hamiltonian operator $\sigma_z$ does \emph{not}
commute with the dissipative operators~$\sigma_+$ or $\sigma_-$. On the other hand, one
may also reformulate the commutation relation \eref{eq:comm1} at the level of the master equation.
Consider a generic one-qubit TLME of \eqnref{eq:mett} with $\Xi_{\omega}(t)\!=\!\mathcal{H}_{\omega} + \mathcal{K}(t)$,
where the $\omega$-encoding coherent contribution reads $\mathcal{H}_{\omega}[\varrho]\!=\! -\ii  \omega  \left[\sigma_z, \varrho\right]/2$
with $\omega\!\neq\!0$, while the second contribution $\mathcal{K}(t)$ (potentially containing both coherent and dissipative
time-dependent parts) is unrestricted yet $\omega$-independent. However, one must impose $\t{Tr}\left\{\Xi_{\omega}(t)[\tau]\right\}\!=\!0$ and
$\Xi_{\omega}[\tau]^{\dag}\!=\!\Xi_{\omega}[\tau^{\dag}]$ for any $\tau \!\in\!\mathcal{L}(\mathbbm{C}^{2^N})$, so that
the resulting dynamical map is trace- and hermiticity-preserving \cite{Gorini1976}.
Using the matrix representation of the two linear maps
$\mathcal{H}_{\omega}$ and $\mathcal{K}(t)$ for each $t$, one concludes that
the TLME generator can be written as in \eqnref{eq:tlg} \emph{if and only if} 
the two contributions commute:
\begin{equation}
[\mathcal{H}_{\omega}, \mathcal{K}(t)] = 0  \qquad \forall t \geq 0.
\label{eq:commme}
\end{equation}
Thus, the commutation property at the level of maps representing dynamics, \eqnref{eq:comm1},
is fully equivalent to the one at the level of maps representing the generators, \eqnref{eq:commme}.

\section{Proof of the ultimate frequency precision bound}
\label{app:proof}
Here we provide a detailed proof of the main result of the paper---%
the ultimate bound on attainable precision dictated by the system short-time behaviour. 
The proof proceeds in three steps. First, we prove the convexity of the
finite-$N$ CE bound with respect to the mixing of channels. Second, we  derive separate bounds
for unital as well as amplitude damping channels. Since every phase covariant map may be decomposed into a mixture of
a unital and an amplitude damping channel, we employ the convexity of the bound to arrive at the universal analytic formula.
Finally, we prove that it is the short-time behaviour of the decoherence parameters that determines the ultimate scaling of the
frequency estimation precision and by considering their explicit short-time expansions
we arrive at \eqnref{eq:step2} of the main text.

\subsection{Finite-$N$ CE bound and its convexity}
\label{app:CEbound_conv}
In order to upper-bound the QFI for a general metrology protocol of \figref{fig:setup}---%
described by IIC dynamics supplemented with ancillary particles,
we use the finite-$N$ CE bound on the QFI introduced in \cite{Fujiwara2008} and developed in \cite{Kolodynski2013,*Kolodynski2014}:
\begin{equation}
\label{eq:boundQFI}
F_\t{Q}\!\left[[\Lambda_{\omega}\ot \mathbbm{1}]^{\ot N}\right] \leq F^\uparrow[\Lambda_{\omega}] = 4  N \min_{\{K_i\}}(\| A \| + (N-1) \|B\|^2),
\end{equation}
where, $F_Q[\Lambda_\omega] = \max_\rho F_Q[\Lambda_\omega(\rho)]$,
the minimisation is performed over equivalent Kraus representations of a given channel $\Lambda_{\omega}$, $\| \cdot \|$ denotes operator norm,
\begin{equation}\label{eq:ab}
A = \sum_{i} \dot{K}^\dagger_i \dot{K}_i, \quad B = \sum_i \dot{K}^\dagger_i K_i
\end{equation}
and $\dot{K}_i = \tfrac{d}{d\omega} K_i$. This approach is similar to \cite{Escher2011}, but has the advantage that one can cast
the problem into a semi-definite program (see \cite{Kolodynski2013,*Kolodynski2014}), and hence obtain a numerical form
of the optimal Kraus representation that yields the tightest bound. Using this as a numerical hint, one can then
proceed analytically with an adequate analytic ansatz for the form of the Kraus representation.
Note that any Kraus representation yields a legitimate bound, so that numerics serve as a helpful tool to simplify the form of the
Kraus representations that need to be considered, without sacrificing neither the tightness nor the validity of the bound.

In what follows, it will prove convenient to think about the minimisation in \eqnref{eq:boundQFI} as a two-step process. First
minimise $\|A\|$, under the constraint of fixed $\| B \| = b$, and then minimise over $b$:
\begin{equation}
\label{eq:boundQFIb}
F^\uparrow[\Lambda_{\omega}] = 4  N \min_b \min_{\{K_i\}, \|B\|=b}(\| A \| + (N-1) \|B\|^2).
\end{equation}

We now prove the convexity of the bound \eref{eq:boundQFI} with respect to mixing of quantum channels, i.e.,
the fact that:

\noindent\emph{%
Given two general maps $\Lambda_{1,\omega}$,
$\Lambda_{2,\omega}$ and
the mixture $\Lambda_\omega=p\Lambda_{1,\omega}+(1-p)\Lambda_{2,\omega}$,
for some probability $0\!\le\!p\!\le\!1$,
one has:}
\begin{equation}
\label{eq:bound_conv}
F^\uparrow[\Lambda_{\omega}] \;\le\;
pF^\uparrow[\Lambda_{1,\omega}]+(1-p)F^\uparrow[\Lambda_{2,\omega}].
\end{equation}
\begin{proof}

The statement \eref{eq:bound_conv} is trivial when considering a single probe evolving through a channel
($N\!=\!1$)---thanks to convexity of the QFI itself \cite{Demkowicz2015,*Toth2014}---but here we want to
demonstrate this property to hold for $N$-probe protocols, which requires some additional arguments.

Let $K_{1,i}$, $i\!\in\!\{1,\dots,n_1\}$, $K_{2,j}$, $j\!\in\!\{1,\dots,n_2\}$ be Kraus representations of
$\Lambda_1$, $\Lambda_2$ respectively (the subfix $\omega$ will be omitted for the sake of simplicity).
We can now easily construct a Kraus representation for the channel $\Lambda = p \Lambda_1 + (1-p) \Lambda_2$
as follows:
\begin{equation}
K_i = \begin{cases}
\sqrt{p} K_{1,i},& i \in \{1,\dots\,n_1\} \\
\sqrt{1-p} K_{2,i-n_1}, & i \in \{n_1+1,n_1+n_2\}.
\end{cases}
\end{equation}
We now utilise \eqnref{eq:boundQFI} and minimise the bound over Kraus representations of the channel.
The latter are generated by
\begin{equation}
\label{eq:kraush}
\tilde{K}_j = \sum_{jj^\prime} \left(e^{- i \mathfrak{h} \omega t}\right)_{jj^\prime} K_{j^\prime},
\end{equation}
where here the hermitian matrix
$\mathfrak{h}$ has dimension $(n_1\!+\!n_2)\!\times\!(n_1\!+\!n_2)$.
By restricting the class of allowed transformation to $\mathfrak{h} = \mathfrak{h}_1 \oplus \mathfrak{h}_2$,
so that Krauses corresponding to $\Lambda_1$ and $\Lambda_2$ are not getting mixed with each other, we can only loosen the bound,
hence we can write:
\begin{multline}
F^\uparrow[\Lambda]  = 4 N \min_b \min_{\mathfrak{h}, \|B\|=b} [\|A\| + (N-1)\|B\|^2]  \leq\\
\leq 4 N  \min_b \min_{\mathfrak{h}=\mathfrak{h}_1\oplus \mathfrak{h}_2, \|B\|=b}\\ \|p A_1 + (1-p) A_2\| +
(N-1)\|p B_1 + (1-p) B_2 \|^2,
\end{multline}
where we have also used the fact that when $\mathfrak{h}_1\oplus \mathfrak{h}_2$, $A$ and $B$ become appropriate weighted sums
of $A_i$, $B_i$. We now make use of convexity of the operator norm to get:
\begin{multline}
F^\uparrow[\Lambda]  \leq 4 N  \min_b \min_{\mathfrak{h}=\mathfrak{h}_1\oplus \mathfrak{h}_2, \|B\|=b}  \\
 p \|A_1\| + (1-p) \|A_2\| + (N-1)(p \|B_1\| + (1-p) \|B_2\|)^2.
\end{multline}
We again restrict the conditions so that instead of requiring $\|B\|\!=\!b$, we impose
stronger constraints $\|B_1\|\!=\!b$, $\|B_2\|\!=\!b$ (Formally,
$\|B_1\|\!=\!b$, $\|B_2\|\!=\!b$ implies $\| B \| \leq b$, and not necessary the equality, but this is exactly what we need here
since $\|A\|$ will then be minimized under in principle even stronger constraint on $\|B\|$, and hence yield a larger value).
Finally,
\begin{multline}
F^\uparrow[\Lambda] \leq 4 N  \bigg[ p \min_b \min_{\mathfrak{h}_1, \|B_1\| = b}\left(\|A_1\| + (N-1) \|B_1\|^2\right) + \\
(1-p) \min_b \min_{\mathfrak{h}_2, \|B_2\| = b}\left(\|A_2\| + (N-1) \|B_2\|^2\right)\bigg] = \\
= p F^\uparrow[\Lambda_1] + (1-p)F^\uparrow[\Lambda_2],
\end{multline}
what completes the proof.
\end{proof}

\subsection{Finite-$N$ CE bound for phase covariant maps}
\label{app:CEbound_cov}

Here, we show how to get a general bound on the extended QFI in the presence
of an IIC channel.
Making use of bound convexity w.r.t.~channels derived in \appref{app:CEbound_conv}, we can get a bound valid for any phase covariant map by
decomposing a general map into a mixture of a unital and an amplitude damping map.
In this section, we use a more compact notation by omitting the map in the argument of $F^{\uparrow}$,
while we explicitly indicate the parameters of the map which fix the form of the upper bound on the extended QFI.
For example, for a generic IIC map, see \eqnref{eq:phasecov_map2}, we write
$F^{\uparrow}[\Lambda_{\eta_\parallel,\eta_\perp, \kappa}]\!\equiv\!F^{\uparrow}_{\eta_{\perp}, \eta_{\parallel}, \kappa}$.
Here, the dependence on $\omega$ is implied, while
$\theta$ can be set to $0$ without loss of generality,
since a rotation around the $\hat{z}$ axis independent from the parameter to be estimated does not modify the (extended) QFI.
Finally, we can also set $\kappa\!\geq\!0$,
since the $z$-axis can always be chosen without loss of generality to
point along the displacement of the phase covariant map.

Let us start with the case of a unital covariant map $\Lambda_{\eta_\parallel,\eta_\perp}$,
which is obtained by setting $\kappa\!=\!0$ in \eqnref{eq:phasecov_map2}.
As stated in \appref{app:CEbound_conv},
in order to evaluate the precision bound based on
the finite-$N$ CE-method \eref{eq:boundQFI}, one must perform
minimisation over all locally inequivalent Kraus representations
of the channel, here the unital phase covariant map.
In general, this corresponds to optimisation over all Kraus operators
that can be genarated from the canonical ones \eqref{eq:Kraus_ops}
via the transformation in \eqnref{eq:kraush},
where $\mathfrak{h}$ is any $4\!\times\!4$ hermitian matrix.
However, by resorting to the SDP-based numerical
analysis introduced in \cite{Kolodynski2013,*Kolodynski2014},
we may state the correct ansatz for the matrix
$\mathfrak{h}$:
\begin{equation}
\mathfrak{h} = \mat{cccc}{
-h & 0 & 0 & 0 \\
0 & h & 0 & 0 \\
0 & 0 &\tfrac{1}{2} & -g \\
0 & 0 & -g & \tfrac{1}{2}}
\end{equation}
with some $h,g\!>\!0$ to be determined.
We find their optimal form by minimising $\|A\|$
in \eqnref{eq:boundQFI} after restricting to
$\mathfrak{h}$ that yield a fixed value of $\|B\|=b t$; where, for later convenience,
we explicitly indicated the linear time dependence of $B$ on time, see \eqnref{eq:ab}.
For a given $b$, we obtain:
\begin{align}
h & = \frac{\eta_\perp^2 - b (1 + \eta_\para)}{1 + \eta_\para - 2 \eta_\perp^2} \\
g & = \frac{\sqrt{\frac{1}{4}(1+\eta_\para)^2-\eta_\perp^2}(1-2b)}{1 + \eta_\para - 2 \eta_\perp^2},
\end{align}
which lead to $A$, $B$ matrices  determining the bound \eref{eq:boundQFI}
that read:
\begin{align}
B &= b t \mat{cc}{-1 & 0 \\ 0 & 1} \\
A &= t^2 \frac{\eta_\perp^2(1 - 4 b) + 2(1+\eta_\para) b^2}{2(1+\eta_\para - 2 \eta_\perp^2)} \mat{cc}{1 & 0 \\ 0 & 1}.
\end{align}
The bound \eqnref{eq:boundQFI} then reads:
\begin{eqnarray}
F^\uparrow_{\eta_\parallel,\eta_\perp}
&=&
4  N t^2 \min_{b} \!\left\{\frac{\eta_\perp^2(1- 4 b) + 2(1+\eta_\para) b^2}{2(1+\eta_\para - 2 \eta_\perp^2)} + (N\!-\!1) b^2\!\right\}
\nonumber \\
&=&
4 N t^2  \min_{b} \!\left\{N b^2 +\frac{\eta_\perp^2(1-2 b)^2}{2 \left(1+\eta_\para-2 \eta_\perp^2\right)}\right\}
\end{eqnarray}
and is minimised for the optimal $b$ value:
\begin{equation}
b= \| B \|/t = \frac{\eta_\perp^2}{N(1+\eta_\para) -2 \eta_\perp^2(N-1)}.
\end{equation}

Finally, we obtain the required general precision bound
for a \emph{unital phase covariant channel} \eref{eq:phasecov_map2} (with $\kappa\!=\!0$):
\begin{eqnarray}
F^{\uparrow}_{\eta_\para,\eta_\perp} =  N t^2 \frac{\frac{2 \eta_\perp^2}{1+\eta_\para - 2 \eta_\perp^2} }{1+ \frac{1}{N} \frac{2 \eta_\perp^2}{1+\eta_\para - 2 \eta_\perp^2}}
= \frac{N^2 t^2}{1+ N \ell},
\label{eq:bounddepoldephas}
\end{eqnarray}
with $\ell = (1+\eta_\parallel - 2 \eta_\perp^2)/2 \eta_\perp^2$.
Note that in the asymptotic limit one gets
\begin{eqnarray}
F^{\uparrow}_{\eta_\para,\eta_\perp}
\underset{N\to\infty}{=}
N t^2 \;\frac{2 \eta_\perp^2}{\left(1+\eta_\para-2 \eta_\perp^2\right)},
\label{eq:bounddepoldephas_as}
\end{eqnarray}
and both the finite-N \eref{eq:bounddepoldephas} and asymptotic \eref{eq:bounddepoldephas_as} bound
are consistent with the ones found previously for
dephasing and depolarisation channels
\cite{Demkowicz2012,Kolodynski2013,*Kolodynski2014}.

Let us now consider the most general case involving the displacement $\kappa\!>\!0$.
Making use of bound convexity w.r.t.~channels derived in \appref{app:CEbound_conv},
we can construct a bound valid for \emph{any} phase covariant map \eref{eq:phasecov_map2}
by decomposing it into ~a \emph{mixture} of a \emph{unital} transformation
and an \emph{amplitude damping} map.
To do so, we must just recall the properties of the amplitude damping map, as
we have already derived a general bound for the unital map case.
General bound for this channel has been derived in \cite{Kolodynski2013,*Kolodynski2014},
but for completeness we re-derive it here.
As noted below \eqnref{eq:cot(theta)}, we can fully parametrise an amplitude
damping channel by the effective displacement parameter $0\!\le\!\kappa\!\le\!1$,
so that the remaining parameters read:~$\eta_\para\!=\!1-\kappa$, $\eta_\perp\!=\!\sqrt{1- \kappa}$,.
As before for the unital case, we deduce from numerics
the optimal ansatz for the matrix $\mathfrak{h}$ to read:
\begin{equation}
\mathfrak{h} = \mat{cccc}{
h & 0 & 0 & 0 \\
0 & 0 & 0 & 0 \\
0 & 0 & g & 0 \\
0 & 0 & 0 & 0},
\end{equation}
where one must set $h\!=\!(1-\kappa-b)(2 - \kappa)/ \kappa$ and $g\!=\!b$
to minimise the finite-$N$ bound \eref{eq:boundQFI} for a given
fixed value of $b=\|B\|/t$
\footnote{This is valid for $b\!\leq\!(1-\kappa)/(2-\kappa)$,
what, however, is always assured for $N\!\geq\!2$ with
$N\!=\!1$ being a special case \cite{Kolodynski2013,*Kolodynski2014}.}.
Plugging in this transformation into \eqnref{eq:boundQFIb} we obtain:
\begin{equation}
F^\uparrow_{\kappa} = 4  N t^2 \min_b \left(\frac{4 -3\kappa}{\kappa} b^2 + \frac{1-\kappa}{\kappa} (1-4 b)+ (N-1)b^2\right),
\end{equation}
which we minimise by optimally setting
\begin{equation}
b = \frac{2(1-\kappa)}{(N-1)\kappa + 4 -3\kappa}
\end{equation}
to obtain, in agreement with \cite{Kolodynski2013,*Kolodynski2014}, the
corresponding precision bounds for an amplitude damping channel:
\begin{eqnarray}\label{eq:boundkappa}
F^{\uparrow}_\kappa
=    N t^2 \frac{\frac{4 (1-\kappa)}{\kappa}}{1+ \frac{1}{N}\frac{4 (1-\kappa)}{\kappa}}
=  \frac{N^2 t^2}{1+ N r},
\end{eqnarray}
with $r = \kappa/(4(1-\kappa))$.
Asymptotically, one obtains
\begin{equation}
F^{\uparrow}_\kappa
\\
 \underset{N\to\infty}{=}
N t^2 \;\frac{4 (1-\kappa )}{\kappa }.
\end{equation}

Finally, let us consider the most general map phase covariant $\Lambda_{\eta_\para,\eta_\perp, \kappa}$
defined via \eqnref{eq:phasecov_map2}
that satisfies the CPTP conditions \eref{eq:cpc} and, without loss of generality, has $\theta=0$ and $k \geq 0$.
Importantly, we may rewrite any such map as the following mixture:
\begin{equation}
\Lambda_{\eta_\para,\eta_\perp, \kappa} = p \Lambda_{\tilde{\eta}_\para,\tilde{\eta}_\perp} + (1-p) \Lambda_{\tilde{\kappa}},
\label{eq:decomp_mix}
\end{equation}
where $p$, $(1-p)$ are the mixing probabilities, $\Lambda_{\tilde{\eta}_\para,\tilde{\eta}_\perp}$
is a valid unital phase covariant map with its corresponding parameters $\tilde{\eta}_\para, \tilde{\eta}_{\perp}$,
and $\Lambda_{\tilde{\kappa}}$ is the amplitude damping map with displacement $\tilde{\kappa}$.
Note that for such decomposition to be valid $0\!\le\!p\!\le\!1$,
$|\tilde\eta_\para|\!\leq\!1$, $|\tilde\eta_\perp| \!\leq\! \tfrac{1}{2}(1+\eta_\para)$
and $0\!\le\!\tilde\kappa\!\le\!1$, the last three conditions ensuring the CPTP of the maps in the mixture.
One may prove that such a decomposition is always possible due to $\eta_{\perp}, \kappa \geq 0$
and the CP constraints in \eref{eq:cpc}. Explicitly,
the decomposition \eref{eq:decomp_mix} may be shown to be valid
after setting the composite channels parameters to:
\begin{eqnarray}
\tilde{\eta}_{\perp} & = & \frac{\eta_{\perp}-(1-p)\sqrt{1-\frac{\kappa}{1-p}}}{p},\nonumber \\
\tilde{\eta}_{\para} & = & \frac{p-1+\eta_{\para}+\kappa}{p},\nonumber \\
\tilde{\kappa} & = & \frac{\kappa}{1-p},
\label{eq:mix_pars}
\end{eqnarray}
where the mixing probability $p$ may be freely chosen to
take any value within the range $\mathcal{P}$ defined such that
\begin{equation}
p\in\mathcal{P}
\;\Leftrightarrow\quad
\mathcal{B}_{+}\,\le\, p\,\le\,
\begin{cases}
1-\kappa & ,\;\eta_{\perp}<\bar{\eta}_{\perp}\\
\mathcal{B}_{-} & ,\;\eta_{\perp}\ge\bar{\eta}_{\perp}
\end{cases},
\label{eq:p_range}
\end{equation}
where
\begin{equation}
\bar{\eta}_{\perp}=\frac{1+\eta_{\para}-\kappa}{2}
\end{equation}
and
\begin{equation}\label{eq:bpm}
\mathcal{B}_{\pm}=\frac{2(1-\kappa)(2+\eta_{\para}\pm2\eta_{\perp})-\!\left(2\eta_{\perp}\pm\eta_{\para}\right)^{2}-(1-\kappa)^{2}}{4(1+\eta_{\para}\pm2\eta_{\perp})}.
\end{equation}

For completeness, let us note that when the original channel
corresponds to an amplitude damping map ($\eta_\para\!=\!1-\kappa$,
$\eta_\perp\!=\!\sqrt{1- \kappa}$), the valid range \eref{eq:p_range}
correctly indicates to choose $p=0$, as then $\eta_{\perp}\!\ge\!\bar{\eta}_{\perp}$
for any $\kappa$ and $\mathcal{B}_\pm\!=\!0$. On the other hand, when a
unital map is considered ($\kappa\!=\!0$), then $\eta_{\perp}\!<\!\bar{\eta}_{\perp}$
by CP \eref{eq:cpc} and $\mathcal{B}_{+}\!=\!\tfrac{1}{4}(3-2\eta_\perp-\eta_\para)$,
so correctly $\mathcal{B}_{+}\!\le\!p\!\le\!1$ and hence by choosing $p=1$
we cancel the contribution from the amplitude damping channel.

Finally, using the convexity property of the bound we are in position to write a
general bound for any phase covariant map as:
\begin{equation}
F^{\uparrow}_{\eta_\para,\eta_\perp,\kappa}
\leq
\min_{p\in\mathcal{P}}
\left\{ p\, F^{\uparrow}_{\tilde{\eta}_\para, \tilde{\eta}_\perp} + (1-p)\, F^\uparrow_{\tilde{\kappa}} \right\},
\label{eq:final_bound}
\end{equation}
where $F^\uparrow_{\tilde{\eta}_\para, \tilde{\eta}_\perp}$, and $F^\uparrow_{\tilde{\kappa}}$ are given
by \eqnsref{eq:bounddepoldephas}{eq:boundkappa} respectively and the parameters of
composite channels $\tilde{\eta}_\para$, $\tilde{\eta}_\perp$, $\tilde{\kappa}$ must be chosen
according to \eqnref{eq:mix_pars} for every $p$ taken from the valid
range $\mathcal{P}$ specified in \eqnref{eq:p_range}.
Lastly, note that $\kappa$ has to be replaced with $|\kappa|$ if it takes on negative values,
see the remark at the beginning of the section.

Finally, let us emphasize that our derived bound \eref{eq:final_bound},
despite potentially not being the tightest one, holds for any $p\!\in\!\mathcal{P}$.
Moreover, we verify numerically that despite exceptions when
considering low $N$ and highly non-unital channels (see for example at the beginning of the next section), $p$ should always be chosen
to take its maximal value within the valid range. This may be explained by
the fact that the CE-based bounds are known to be tighter for unital channels
\cite{Demkowicz2012,Kolodynski2013,*Kolodynski2014},
so that their contribution should be intuitively maximised
when decomposing into the mixture of \eqnref{eq:decomp_mix}.
We thus conclude that one can approximate the optimal
mixing probability as
\begin{equation}
p_\t{opt}\,\approx\,\label{eq:pot}
\begin{cases}
1-\kappa & ,\;\eta_{\perp}<\bar{\eta}_{\perp}\\
\mathcal{B}_{-} & ,\;\eta_{\perp}\ge\bar{\eta}_{\perp}
\end{cases},
\end{equation}
which we then use when employing bound \eref{eq:final_bound},
unless otherwise stated.

\subsection{The ultimate bound on precision
for general phase covariant dynamics}
\label{app:Zeno_bound}

Now we are in the position to prove the validity of the general limit for the frequency estimation
under any IIC dynamics presented in \eqnsref{eq:step2}{eq:fincases2} of the main text.
First, we prove that the optimal interrogation time (duration of each experimental shot) lies
asymptotically in the short time regime, i.e.,~$t_\t{opt}(N)\!\to\!0$ with $N$ as $N^{-a}$ for some power $a$,
unless the dynamics trivially becomes fully decoherence-free for some finite $t$. Hence, apart from such
(unrealistic) case, the precision is always asymptotically limited to follow the SQL-like scaling,
unless the interrogation time is cunningly chosen to be vanishing with $N$.

Consider an IIC map \eref{eq:phasecov_map2} with  $\eta_\perp\!<\!1$ for a given fixed time $t$.
Then, stemming from \eqnref{eq:boundQFI} and the results of previous section,
one can construct an \emph{asymptotic} bound
\begin{equation}\label{eq:prr}
\lim_{N \rightarrow \infty} \frac{F_\t{Q}\!\left[[\Lambda_{\omega} \ot \mathbbm{1}]^{\ot N}\right]}{N}
\leq   t^2 c,
\end{equation}
with $c\!>\!0$ being some finite constant, so that no super-classical scaling is indeed possible
unless $t\!\to\!0$.

The factor $c$ follows directly from \eqnref{eq:final_bound} after substituting adequately
for the unital and amplitude damping channels bounds of \eqnsref{eq:bounddepoldephas}{eq:boundkappa}.
In case $r\!\neq\!0$ and $\ell\!\neq\!0$ it simply reads
(without the minimisation over $p$ for convenience)
\begin{equation}\label{eq:prr2}
c = \frac{  p r + (1-p) \ell }{\ell r}.
\end{equation}
If $\ell \!=\! 0$, but also $p \!=\! 0$,
one is left with the amplitude damping channel only, so that for $r \neq 0$, \eqnref{eq:prr} holds with $c \!=\! 1/r$.
On the other hand, $r \!=\! 0$ only if $\kappa \!=\! 0$, i.e $\Lambda_{\omega}$ is a unital map and its QFI is thus bounded by $N^2 t^2/(1+N \ell)$, see \eqnref{eq:bounddepoldephas}:~
for $\ell \neq 0$, one gets \eqnref{eq:prr} with $c \!=\! 1/\ell$.
In other words, one cannot have super-classical scaling, unless
$\ell \!=\! 0$ and $p \neq 0$ or
$r \!=\! \ell \!=\! 0$.
But we now show that these conditions are excluded
by the hypothesis $\eta_{\perp} \!\!<\!\! 1$.
First, $\ell\!=\! 0$ iff $\tilde{n}_{\perp}^2 \!=\!(1+\tilde{\eta}_{\parallel})/2$,
but since for the CPTP of the unital covariant map we know that, see \eqnref{eq:cpc}, $\tilde{n}_{\perp} \!\leq\!(1+\tilde{\eta}_{\parallel})/2$
and $|\tilde{\eta}_{\parallel}| \!\leq\! 1$, it follows that $\ell \!=\! 0$ iff $\tilde{\eta}_{\parallel} \!=\! \tilde{\eta}_{\perp} \!=\! 1$. Indeed, $\tilde{\eta}_{\parallel} \!=\! 1$ is equivalent to $\eta_{\parallel} + |\kappa| \!=\!1$,
see \eqnref{eq:mix_pars} (recalling that $\kappa$ has to be changed to $-\kappa$ if it takes a negative value).
Now, let us set for convenience $p \!=\! \mathcal{B}_{+}$, see \eqnref{eq:bpm}
($p\!=\!\mathcal{B}_{-}$ or $p \!=\! 1-|\kappa|$ would not be the optimal choice in this case). Then, one can show, also using $\eta_{\parallel} + |\kappa| \!=\! 1$,
that $\tilde{\eta}_{\perp}\!=\!1$ iff $\eta_{\perp} \!=\! 0$
or $p \!=\! 0$. The former case can be excluded since it corresponds to QFI equal to 0.
So, we are left only with the case where both $\ell$ and $r$ are equal to zero, which corresponds to $\kappa \!=\! 0$ and $\eta_{\parallel} \!=\!\eta_{\perp}
 \!=\!1$---the decoherence-free case excluded by the "no full-revival" assumption.

Now, given the full IIC dynamics,
$\left\{[\Lambda_{\omega}(t) \ot \mathbbm{1}]^{\ot N}\right\}_{t\geq0}$, such that $\eta_{\perp}(t)\!<\!1$ for all $t\!>\!0$,
the previous result implies that the optimal evaluation time lies in the short-time regime,
which, along with the general bound derived in the previous paragraph, will allow us to get the precision
limit quoted in \eqnsref{eq:step2}{eq:fincases2} of the main text.
The quantum Cram{\'e}r-Rao bound (QCRB---\eqnref{eq:una} of the main text) further
lower-limited with use of \eqnref{eq:final_bound} yields
\begin{equation}
\label{eq:crbmixture}
\Delta^2 \omega_N\,T \geq \min_{t} \frac{1}{N^2 t} \left(\frac{p(t)}{1+N \tilde{\ell}(t)} + \frac{1-p(t)}{1+ N \tilde{r}(t)}\right)^{-1},
\end{equation}
where
\begin{eqnarray}
\tilde{\ell}(t) &=& \frac{1-2\tilde{\eta}_{\perp}(t)^2+\tilde{\eta}_{\parallel}(t)}{2 \tilde{\eta}_{\perp}(t)^2}, \label{eq:ell}\\
\tilde{r}(t) &=&\frac{\tilde{\kappa}(t)}{4(1-\tilde{\kappa}(t))}, \label{eq:err}
\end{eqnarray}
and the tilde parameters are, for any fixed $t$, as in \eqnref{eq:mix_pars}.
We are looking for the optimal evaluation time
$t_{\t{opt}}(N)$ where the right-hand side (r.h.s.)~of \eqnref{eq:crbmixture} attains its minimum value.
However, for any sequence of interrogation time settings with $N$, $t(N)$,
such that $t(N)\!\rightarrow\!t_{\infty}$  as $N\!\rightarrow\!\infty$, with $0\!<\! t_{\infty}\!\leq\!T$,
using \eqnref{eq:prr} ($\eta_{\perp}(t_{\infty})\!<\!1$ by hypothesis), we get
\begin{equation}\label{eq:finn}
\lim_{N\rightarrow \infty} \frac{\Delta^2 \omega_N\,T }{N^{-1}} \geq \frac{1}{t_{\infty} c(t_{\infty})}:
\end{equation}
the precision is asymptotically limited by the SQL.

Thus, to go beyond the SQL we have to look for a minimum of the r.h.s. of \eqnref{eq:crbmixture} which goes to 0 for $N\!\rightarrow\!\infty$.
We show that this is always possible for times short enough
by using the short-time expansion of the dynamical parameters
(\eqnref{eq:cov_coeffs} of the main text):
\begin{eqnarray}
\eta_{\perp}(t) &=& 1-\alpha_{\perp} t^{\beta_{\perp}}+ o(t^{\beta_{\perp}}) \nonumber\\
\eta_{\parallel}(t) &=& 1-\alpha_{\parallel} t^{\beta_{\parallel}}+o( t^{\beta_{\parallel}}) \nonumber\\
\kappa(t) &=& \alpha_{\kappa} t^{\beta_{\kappa}}+o(t^{\beta_{\kappa}}),
\label{eq:coeffs_app}
\end{eqnarray}
with $\alpha_{\perp}, \alpha_{\parallel}\!\geq\!0$ and $\alpha_{\perp}\!\neq\!0$
(since we assume $\eta_\perp(t)\!<\!1$ for $t\!>\!0$),
and $\beta_{\perp}, \beta_{\parallel}, \beta_\kappa \geq 1$
(since we assume the dynamical parameters to be
functions of class $C^1(\mathbbm{R}_0^+)$).
The CPTP conditions in \eqnref{eq:cpc} lead
to the following constraints:
\begin{equation}
 \left\{\begin{array}{cc}
\beta_{\perp} \leq \beta_{\parallel} &  \t{and} \,\, \alpha_{\parallel} \leq 2 \alpha_{\perp} \,\,{\t{if}}\,\,\beta_{\perp} = \beta_{\parallel} \\
\beta_{\parallel} \leq \beta_{\kappa} & \t{and} \,\, |\alpha_{\kappa}| \leq \alpha_{\parallel} \,\,{\t{if}}\,\,\beta_{\parallel} = \beta_{\kappa}.
\end{array}\label{eq:cpcst}
     \right.
\end{equation}
Moreover, since we use $p(t)$ as in
\eqnref{eq:pot}, its expansion will depend on the relation between $\eta_{\perp}(t)$ and $\bar{\eta}_{\perp}(t)$.
It is thus convenient to express the latter in terms of the coefficients and powers in the
expansions of the parameters $\eta_{\perp}(t), \eta_{\parallel}(t), \kappa(t)$. Within the constraints set by \eqnref{eq:cpcst}, one has $\eta_{\perp}(t)<\bar{\eta}_{\perp}(t)$
for short times if
\begin{equation}
 \left\{\begin{array}{cc}
\beta_{\perp} < \beta_{\parallel} \\
\beta_{\perp} = \beta_{\parallel} < \beta_{\kappa} \,\,\,\t{and} \,\,\, \alpha_{\parallel} \neq 2 \alpha_{\perp} \\
\beta_{\perp} = \beta_{\parallel} = \beta_{\kappa}  \,\,\, \t{and} \,\,\,  \alpha_{\perp}-\frac{\alpha_{\parallel}}{2} > \frac{\alpha_{\kappa}}{2}.
\end{array}\label{eq:cass}
     \right.
\end{equation}

Let us assume for the moment
that if $\beta_{\perp} = \beta_{\parallel} \leq \beta_{\kappa}$ then $\alpha_{\parallel} \neq 2\alpha_{\perp}$.
The short time expansion of $p(t)$ hence reads
\begin{equation}
p(t) = \,\label{eq:pot2}
\begin{cases}
1-|\alpha_{\kappa}| t^{\beta_k} +o(t^{\beta_k}) &\\
1-\frac{1}{2}\alpha t^{\beta_k}
-\frac{1}{2}\alpha_{\perp} t^{\beta_{\perp}}+\frac{1}{4}\alpha_{\parallel}t^{\beta_{\parallel}} + o(t^{\beta_\perp}),
\end{cases}
\end{equation}
with $\alpha \!=\! |\alpha_{\kappa}|+\alpha_{\kappa}^2/(4 \alpha_{\perp}-2 \alpha_{\parallel})$,
where the two cases refer to, respectively,  $\eta_{\perp}(t)<\bar{\eta}_{\perp}(t)$ and $\eta_{\perp}(t)\geq\bar{\eta}_{\perp}(t)$.
In any case, we can write $p(t) \!=\! 1-\alpha_p t^{\beta_p}$, where $\alpha_p > 0$ and $\beta_p\geq 1$ are fixed by \eqnref{eq:pot2}.
Thus, we are in position to derive the short-time
expansions of $\tilde{\eta}_{\parallel}(t), \tilde{\eta}_{\perp}(t)$ and $\tilde{\kappa}(t)$,
defined in \eqnref{eq:mix_pars}, from which we can then write the expansions
of the bound coefficients \eref{eq:ell} and \eref{eq:err}:
\begin{eqnarray}
\tilde{\ell}(t)&=& 2 \alpha_{\perp} t^{\beta_{\perp}}-\frac{1}{2} |\alpha_{\kappa}| t^{\beta_k}-\frac{1}{2} \alpha_{\parallel} t^{\beta_{\parallel}} + o(t^{\beta_k}) \nonumber\\
\tilde{r}(t)&=& \frac{|\alpha_{\kappa}|}{4 \alpha_p} t^{\beta_\kappa-\beta_p} + o( t^{\beta_k-\beta_p} ).\label{eq:explr}
\end{eqnarray}
We write for compactness $\tilde{\ell}(t) \!=\! \alpha_{\tilde{\ell}} t^{\beta_{\tilde{\ell}}}+ o(t^{\beta_{\tilde{\ell}}})$ ($\tilde{\ell}(0) \!=\! 0$)
and $\tilde{r}(t) \!=\! \tilde{r}_0 + \alpha_{\tilde{r}} t^{\beta_{\tilde{r}}}$,
where $\tilde{r}_0 \equiv \tilde{r}(0)$ is different from 0 if $\beta_p \!=\! \beta_\kappa$).

We may now substitute the above-derived expansions
to derive short time expression for the precision bound \eref{eq:crbmixture}:
\begin{eqnarray}\label{eq:prev}
&&\lim_{t\rightarrow 0^+}\Delta^2 \omega_N\,T \geq \nonumber \\ && \lim_{t\rightarrow 0^+}
\frac{(1+ N(r_0+\alpha_{\tilde{r}} t^{\beta_{\tilde{r}}}))(1+N \alpha_{\tilde{\ell}} t^{\beta_{\tilde{\ell}}})}
{N^2 t \left(1+N(r_0+\alpha_{\tilde{r}} t^{\beta_{\tilde{r}}}+\alpha_p \alpha_{\tilde{\ell}} t^{\beta_p+\beta_{\tilde{\ell}}})\right)}.
\end{eqnarray}
For all the cases in \eqnref{eq:cass},
one has $\beta_p \!=\! \beta_\kappa$, so that $\tilde{r}_0 \neq 0$ if $\alpha_\kappa \neq 0$,
while for $\beta_{\perp} \!=\! \beta_{\parallel} \!=\! \beta_{\kappa}$
one has $\beta_p + \beta_{\tilde{\ell}} > \beta_r$, see \eqnsref{eq:pot2}{eq:explr}.
In both situations \eqnref{eq:prev} reduces to
\begin{equation}\label{eq:lim2}
\lim_{t\rightarrow 0^+}\Delta^2 \omega_N\,T \geq  \lim_{t\rightarrow 0^+} \frac{1+N \alpha_{\tilde{\ell}} t^{\beta_{\tilde{\ell}}}}{N^2 t},
\end{equation}
i.e., the only relevant contribution to the bound is that coming from the unital part of the mixture.
Actually, \eqnref{eq:lim2} holds also for the other possible CPTP dynamics
specified in \eqnref{eq:cpcst}.
Indeed, this is the case if $\alpha_\kappa\!=\!0$, i.e., $\kappa \!=\! 0$, whereas for $\beta_{\perp} \!=\! \beta_{\parallel} \!\leq\! \beta_{\kappa}$ and $\alpha_{\parallel} \!=\! 2 \alpha_{\perp}$, it can be shown proceeding as above, but taking into account
the higher order terms in the expansion of the parameters in $p(t)$, see \eqnref{eq:bpm}.

Now, the function $(1+ N \alpha_{\tilde{\ell}} t^{\beta_{\tilde{\ell}}})/(N^2 t)$ has a local minimum at
\begin{equation}\label{eq:bart}
\bar{t}(N) = \frac{1}{\left(\alpha_{\tilde{\ell}} N (\beta_{\tilde{\ell}} - 1)\right)^{1/\beta_{\tilde{\ell}}}},
\end{equation}
which goes to 0 as $1/N^{1/\beta_{\tilde{\ell}}}$ for $N\rightarrow\infty$
and yields
\begin{equation}
\label{eq:last}
\lim_{N \rightarrow \infty} \frac{\Delta^2 \omega_N\,T}{N^{-(2 \beta_{\tilde{\ell}}-1)/\beta_{\tilde{\ell}}}} \geq
\frac{\alpha_{\tilde{\ell}}^{1/\beta_{\tilde{\ell}}} \beta_{\tilde{\ell}}}{(\beta_{\tilde{\ell}}-1)^{(\beta_{\tilde{\ell}}-1)/\beta_{\tilde{\ell}}}}.
\end{equation}
It is thus clear that for $\beta_{\tilde{\ell}}\!>\!1$ the precision estimation can overcome the SQL scaling.

Before writing explicitly $\beta_{\tilde{\ell}}$ and $\alpha_{\tilde{\ell}}$, let us note that
if $\beta_{\perp} \!=\! \beta_{\parallel} \!=\! \beta_{\kappa}$
and $\alpha_{\parallel} \!=\!|\alpha_{\kappa}|\!=\!2 \alpha_{\perp}$, one would need to
take into account higher order terms in the expansion of $\tilde{\ell}(t)$, see \eqnref{eq:explr}.
However, we can take advantage once again of the full range $\mathcal{P}$
of values of $p$ allowing for a meaningful mixture, see \appref{app:CEbound_cov},
and set $p(t) \!=\! \mathcal{B}_+(t)$---instead of using \eqnref{eq:pot} yielding \eqnref{eq:pot2}.
Now, $p(0) \!=\! 0$, $r(t) \!=\! |\alpha_{\kappa}| t^{\beta_\perp}/4$ and
one gets Eqs.~\eqref{eq:lim2}-\eqref{eq:last} with $\alpha_{\tilde{r}}$ and $\beta_{\tilde{r}}$ instead of, respectively, $\alpha_{\tilde{\ell}}$ and $\beta_{\tilde{\ell}}$.
Indeed, also for $\beta_{\perp} \!=\! \beta_{\parallel} \!=\! \beta_{\kappa}$
and $\alpha_{\parallel}/2 + |\alpha_{\kappa}|/2-2 \alpha_{\perp}\neq0$ one could get the same bound, so that
in this case
we can maximise between this result and that obtained by using $p(t)$ as in \eqnref{eq:pot}.

Referring to the short-time expansion of $\tilde{\ell}(t)$ in \eqnref{eq:explr} (and $\tilde{r}(t)$ if $\beta_{\perp} \!=\! \beta_{\parallel} \!=\! \beta_{\kappa}$)
and to the different cases allowed by CPTP constraints of \eqnref{eq:cpcst},
we immediately see that one always has $\beta_{\tilde{\ell}} \!=\! \beta_{\perp}$.
As said, the scaling of the asymptotic precision is fixed by the short time expansion of $\eta_{\perp}$, i.e., \eqnref{eq:last}
provides us with \eqnref{eq:step2} of the main text, where the constant $D$ depends instead on the
expansion of all the parameters. Using \eqnsref{eq:explr}{eq:last} (with $\tilde{\ell}$ possibly replaced by $\tilde{r}$ in the the latter
when $\beta_{\perp} \!=\! \beta_{\parallel} \!=\! \beta_{\kappa}$),
we end up with the claimed lower bound, i.e., \eqnref{eq:step2} of the main text, where
\begin{equation}\label{eq:fincas}
D = \frac{\alpha^{1/\beta_{\perp}} \beta_{\perp}}{(\beta_{\perp}-1)^{(\beta_{\perp}-1)/\beta_{\perp}}},
\end{equation}
and
\begin{equation}
\alpha =  \left\{\begin{array}{cc}
2 \alpha_{\perp} &\quad \beta_{\perp}< \beta_{\parallel}; \\
2 \alpha_{\perp}-\frac{\alpha_{\parallel}}{2} &\quad  \beta_{\perp}= \beta_{\parallel}<\beta_\kappa; \\
\max\left\{2 \alpha_{\perp}-\frac{\alpha_{\parallel}}{2} -\frac{|\alpha_{\kappa}|}{2}, \frac{|\alpha_{\kappa}|}{4}\right\}&\quad  \beta_{\perp}= \beta_{\parallel}= \beta_k.
\end{array}\label{eq:fincases}
     \right.
\end{equation}

Finally, note that for $\beta_\perp \!=\!1$ both the bound for finite times and for $t\!\rightarrow\!0$ give
a $1/N$ scaling of the estimation error, see \eqnsref{eq:finn}{eq:last}. In this case,
\eqnref{eq:step2} of the main text holds, but the minimum value of the constant $D$
has to be determined by a comparison between the value corresponding to local minima 
at finite $t$, as in \eqnref{eq:finn}, and the one adequate for the $t\!\to\!0$ limit,
specified by \eqnsref{eq:last}{eq:fincas}.

\section{Consequences of the Zeno regime exhibited at short timescales}
\label{app:driving_H}
Here, we show that a quadratic decay of the survival probability on the short time, i.e. the so-called Zeno regime
\cite{Facchi2008,*Pascazio2014}, fixes the power in the expansion of the parameters of the
dynamical map $\Lambda_{\omega}(t)$
and, in particular, it leads to $\beta_{\perp}\!\!=\!\!2$, so that from \eqnref{eq:last} one recovers
the ultimate $N^{-3/2}$ scaling.

Let us consider a reduced evolution
which is obtained by taking exactly the partial trace over a global unitary dynamics ruled by a possibly time-dependent Hamiltonian $H(t)$
and for an initial product state $\rho\!=\! \rho_S\ot\ketbra{\psi_E}{\psi_E}$, i.e.:
\begin{equation}
\Lambda(t)\left[\rho_S\right] = \t{tr}_E\left\{U(t)[\rho_S \ot \ketbra{\psi_E}{\psi_E}] U^{\dag}(t)\right\}.
\end{equation}
It is then easy to see that the quadratic decay of the survival probability at the level of the overall
unitary dynamics, $|\braket{\psi}{U(t) |\psi}|^2 \!=\! 1- \lambda t^2 + o(t^2)$ for any $\ket{\psi}$, implies an analogous decay on the survival probabilities of the open system's state.
Given any initial pure state $\ket{\psi_S}$
(and denoting as $\ket{\psi} \!=\! \ket{\psi_S}\ot \ket{\psi_E}$ the corresponding initial overall
system+environment state), one has
\begin{equation}\label{eq:zeno}
\bra{\psi_S} \Lambda(t)\left[\ketbra{\psi_S}{\psi_S}\right] \ket{\psi_S} = 1 - \lambda_S t^2 + o(t^2),
\end{equation}
where
\begin{equation}
\lambda_S = \bra{\psi}H(0)^2\ket{\psi} - \bra{\psi_S}\t{tr}_E\left\{ H(0)[\ketbra{\psi}{\psi}] H(0) \right\}\ket{\psi_S}.
\end{equation}
For our purposes, it is convenient to express this relation in terms of the Bloch vector ${\bf v}$ associated with $\ket{\psi_S}$.
Given two states $\rho$ and $\tau$ with Bloch vectors ${\bf v}_{\rho}$ and ${\bf v}_{\tau}$, one has, see \eqnref{eq:blsp},
\begin{equation}
\t{Tr}\left\{ \rho \tau\right\} = \frac{1}{2}(1 +  {\bf v}_{\rho} \cdot {\bf v}_{\tau}),
\end{equation}
so that given a map $\Lambda_{\omega}(t)$ as in \eqnref{eq:phasecov_map}, we get
\begin{eqnarray}
&&\bra{\psi_S} \Lambda_{\omega}(t)[\ketbra{\psi_S}{\psi_S}] \ket{\psi_S}
=\frac{1}{2}\left(1+ v_z \kappa(t) + v^2_z \eta_{\parallel}(t) \right. \nonumber\\
&&\left.+ (v^2_x+v^2_y) \eta_{\perp}(t) \cos\phi(t) \right).\nonumber
\end{eqnarray}
If we now focus on the initial pure state such that $v_z=1$, so that $v_x=v_y = 0$,
we get
\begin{equation}
\bra{\psi_S} \Lambda_{\omega}(t)[\ketbra{\psi_S}{\psi_S}] \ket{\psi_S}  = \frac{1}{2}\left(1+\kappa(t) + \eta_{\parallel}(t)\right),
\end{equation}
and hence, by using the short time expansion of $\kappa(t)$ and  $\eta_{\parallel}(t)$, \eqnref{eq:zeno} implies $\beta_{\parallel} = 2$;
recall that $\beta_{\parallel} \leq \beta_k$ for CPTP constraints in \eqnref{eq:cpc},
which thus also imply
$\beta_{\perp} \leq 2$.
Now, instead, consider the case $v_z=0$,  $v_x^2+v_y^2 = 1$, so that
\begin{equation}\label{eq:notb}
\bra{\psi_S} \Lambda_{\omega}(t)[\ketbra{\psi_S}{\psi_S}] \ket{\psi_S}  = \frac{1}{2}\left[1+ \eta_{\perp}(t) \cos\phi(t) \right].
\end{equation}
But then, using the short-time expansion of the coefficients and since $\beta_{\perp} \leq 2$, it is easy to see that
\eqnsref{eq:zeno}{eq:notb} imply
$\beta_{\perp} = 2$ for any $\theta(t) \in C^1(\mathbbm{R}_0^+)$.

Finally, let us emphasize that
in several situations of interest the survival probability does not decay quadratically on short times.
This is of course the case when one deals with effective descriptions, such as the semigroup one, which involve a
coarse graining in time \cite{Breuer2002}.
Moreover, also at the level of the global unitary dynamics, it may happen that the survival probability is not an analytic function of time at $t = 0$;
this situation was investigated for example in \cite{Antoniou2001}, where a decay of $|\braket{\psi(t)}{\psi}|^2$ with the power of $3/2$ was found.

\section{Performance of GHZ states}
\label{app:GHZ}

We consider a standard quantum metrology protocol that employs solely probes
prepared in a GHZ state without utilising the ancillary qubits \cite{Huelga1997,Matsuzaki2011,Chin2012}.
Then, the general form of the system state at time $t$, see \eqnref{eq:fin_state} of the main text reads
\begin{equation}
\rho_\omega(t) = \Lambda_{\omega}(t)^{\ot N} [\psi_\t{\tiny GHZ}^N] \ot \sigma^{N_\t{A}},
\end{equation}
where $\ket{\psi_\t{\tiny GHZ}^N}\!=\!\frac{1}{\sqrt{2}}(\ket{0}^{\ot N}\!+\!\ket{1}^{\ot N})$ and
$\sigma^{N_\t{A}}$ are respectively the $N$-particle GHZ and the irrelevant ancillary states.

Crucially, the QFI of the output state calculated w.r.t.~the frequency $\omega$ can
then be evaluated only considering the probes:~%
$F^{N}_{\t{\tiny GHZ}}\!=\!
F_\t{Q}[\rho_{\omega}(t)]\!=\!
F_\t{Q}[\Lambda_{\omega}(t)^{\ot N}[\psi_\t{\tiny GHZ}^N]\!\ot\! \sigma^{N_\t{A}}]\!=\!
F_\t{Q}[\Lambda_{\omega}(t)^{\ot N}[\psi_\t{\tiny GHZ}^N]]$.
Moreover, for the general phase-coviant noise, i.e.~the map specified
by \eqnref{eq:phasecov_map2}, it reads
\begin{equation}
F^{N}_{\t{\tiny GHZ}}
=
\frac{t^2 N^2 \eta_\perp^{2 N}}{2^{-(N+1)} \left(A_{-,-}^N+A_{+,-}^N+A_{-,+}^N+A_{+,+}^N\right)}
\label{eq:GHZ}
\end{equation}
with $A_{\pm,\pm}=1\pm\eta_\parallel\pm\kappa$.

In order to obtain the best performance using GHZ states, we need to find the optimal interrogation time
as a function of $N$, so that the QCRB (the r.h.s.~of \eqnref{eq:una} in the main text) is minimised. 
Equivalently, this corresponds to
computing $\max_t F^N_\t{\tiny GHZ}/t$. We are interested in the asymptotic performance for large $N$
and hence, as argued in the main text, we may focus on short timescales of the evolution.
Plugging into \eqref{eq:GHZ} the short-time expansions of $\eta_\perp$, $\eta_\parallel$ and $\kappa$
specified in \eqnref{eq:coeffs_app}, we obtain the asymptotic expression:
\begin{multline}
\label{eq:GHZa}
F^{N}_{\t{\tiny GHZ}}/t \;\overset{N \rightarrow \infty}{=} \\
\frac{t N^2 (1- \alpha_\perp t^{\beta_\perp})^{2N}}{
2^{-(N+1)}[(2-\alpha_\parallel t^{\beta_\parallel} + \alpha_\kappa t^{\beta_\kappa})^N   + (2-\alpha_\parallel t^{\beta_\parallel} - \alpha_\kappa t^{\beta_\kappa})^N ]
}.
\end{multline}
Consider the general asymptotic expression for
the optimal interrogation time for the GHZ-based strategy:
\begin{equation}
\label{eq:tN}
t_\t{\tiny GHZ}(N) \overset{N \rightarrow \infty}{=}
\frac{1}{(\alpha_t N)^{1/\beta_t}}.
\end{equation}
It is clear that it is the enumerator in \eqnref{eq:GHZa} that is responsible for the resulting
precision scaling with $N$.
In particular, substituting for $t$ according to \eqnref{eq:tN} it reads
\begin{equation}
\alpha_t^{-1/\beta_t} N^{2-1/\beta_t}(1-\alpha_\perp (\alpha_t N)^{-\beta_{\perp}/\beta_t})^{2N}.
\end{equation}
Hence, in order to attain the best scaling, we focus on the term in front of the parenthesis above
and set $\beta_t$ as large as possible.
However, if we take $\beta_t \!>\!\beta_\perp$, the term in the parenthesis
leads to an exponential decay with $N$, whereas while taking $\beta_t\!<\!\beta_\perp$
the exponentiation of the parenthesis term asymptotically yields $1$.
Thus, we must optimally set $\beta_t \!=\! \beta_\perp$.

Bearing in mind that the CP condition of the phase covariant map
implies $\beta_\perp \!\leq\! \beta_\parallel \!\leq\! \beta_\kappa$,
we assume for the moment that $\beta_\perp \!=\! \beta_\parallel \!=\! \beta_\kappa$,
for which \eqnref{eq:GHZa} becomes:
\begin{equation}
F^{N}_{\t{\tiny GHZ}}/t
\; \overset{N \rightarrow \infty}{=}\;
\frac{\alpha_t^{-1/\beta_\perp} \ee^{-2 \alpha_\perp/\alpha_t}}
{\ee^{-\alpha_\parallel/2 \alpha_t} \cosh \frac{\alpha_\kappa}{2 \alpha_t}}
N^{2 - 1/\beta_\perp}.
\label{eq:Fghz_as}
\end{equation}
\eqnref{eq:Fghz_as} is maximised by choosing $\alpha_t$ that
satisfies the transcendental equation:
\begin{equation}
 \frac{\alpha_t}{\beta_\perp} = \frac{\alpha_\kappa}{2} \tanh \frac{\alpha_\kappa}{2 \alpha_t}  + 2 \alpha_\perp - \frac{\alpha_\parallel}{2},
\label{eq:alphat_opt}
\end{equation}
which can be always solved numerically.
Note that, since $\alpha_t\!>\!0$ and $\tanh$ for positive (negative) arguments
takes values in $(0,1]$ ($[-1,0)$), the solution of \eqnref{eq:alphat_opt} will
always yield
\begin{equation}
\frac{\alpha_t}{\beta_\perp}
\in
\left[2\alpha_\perp - \frac{\alpha_\parallel}{2}, 2\alpha_\perp-\frac{\alpha_\parallel}{2} + \frac{|\alpha_\kappa|}{2}\right].
\label{eq:alphatrange}
\end{equation}
In case of $\beta_\perp \!<\! \beta_\kappa$, the $\cosh$ term in \eqnref{eq:Fghz_as}
is asymptotically irrelevant and should be dropped, what corresponds to simply setting
$\alpha_\kappa\!=\!0$ in the above formulae. Similarly, it is enough to set $\alpha_\parallel\!=\!0$
whenever $\beta_\perp\!<\!\beta_\parallel$.

Resorting to \appref{app:Zeno_bound} and the constraints imposed on the
short-time coefficients by the CPTP condition, one may show that for any phase covariant
noise (not exhibiting ``full revival'') there always exist an optimal, non-zero
$\alpha_t$ in the valid range \eref{eq:alphatrange}. Hence, \eqnref{eq:Fghz_as}
implies that the GHZ-based strategy always allows for $F^{N}_{\t{\tiny GHZ}}/t\!\overset{N \rightarrow \infty}{\sim}\!N^{2 - 1/\beta_\perp}$,
so that the asymptotic precision scaling $\Delta^2\omega_N\!\sim\!1/N^{(2 \beta_\perp-1)/\beta_\perp}$
indicated by \eqnref{eq:last} (and \eqnref{eq:step2}) is indeed always attainable.

Specifically, whenever $\beta_\perp \!<\! \beta_\kappa$,
or for the special case of unital IIC dynamics (when $\kappa\!=\!0$
and thus by definition $\alpha_\kappa\!=\!0$), the solution of
\eqnref{eq:alphat_opt} reads:
\begin{equation}
\frac{\alpha_t}{\beta_\perp}
\;\underset{(\alpha_\kappa=0)}{=}\;
\begin{cases}
2 \alpha_\perp, &  \beta_\perp < \beta_\parallel \\
2 \alpha_\perp - \frac{\alpha_\parallel}{2}, & \beta_\perp = \beta_\parallel
\end{cases}.
\label{eq:dupa}
\end{equation}
Therefore, in agreement with the result stated in \eqnref{eq:GHZ_as_const} of the main text,
for $\kappa\!=\!0$
\begin{equation}
F^{N}_{\t{\tiny GHZ}}/t
\quad\underset{(\kappa=0)}{\overset{N \rightarrow \infty}{=}}\quad
\frac{N^{2-1/\beta_\perp}}{[\alpha_l \beta_\perp\ee]^{1/\beta_\perp}},
\end{equation}
with $\alpha_l\!=\!\alpha_t/\beta_\perp$ defined via \eqnref{eq:dupa} (similarly to \eqnref{eq:ellell}).

\section{Shabani-Lidar post-Markovian noise model}
\label{app:SLmodel}

The model of \citet{Shabani2005} (SL) has been grounded phenomenologically
to describe qubit evolution that interpolates between the exact Kraus map and
semigroup dynamics and it has been later shown
to provide a broad platform to study non-semigroup and non-Markovian
evolutions~\cite{Maniscalco2006,*Maniscalco2007,*Mazzola2010}.

In particular, given a spin $1/2$ particle interacting
with a bosonic reservoir under rotating wave approximation,
the model can be also formulated in terms of a TLME with generator 
of the form in \eref{eq:tlg}. Defining
\begin{equation}
R=\frac{\gamma_0}{\gamma}(2n +1)
\quad\t{and}\quad
f(R,t)=\frac{1-R}{1-\ee^{-(1-R)t}R},
\end{equation}
one finds the
time-dependent decay rates of \eqnref{eq:see} to read:
\begin{eqnarray}
\label{eq:gammas_SLmodel}
\gamma_{+}(t) & = & \gamma\frac{n}{2n+1}\left[1-f(R,\gamma t)\right],\\
\gamma_{-}(t) & = & \gamma\frac{n+1}{2n+1}\left[1-f(R,\gamma t)\right],\nonumber \\
\gamma_{z}(t) & = & \frac{\gamma}{4}\left[1-2\,f\!\left(\frac{R}{2},\gamma t\right)+f(R,\gamma t)\right].\nonumber
\end{eqnarray}
Here, ~$\gamma_0$ is the phenomenological dissipation constant, $n$
the mean number of excitations in the reservoir, and $\gamma$ the effective memory rate.
The resulting master equation yields always a well-defined CPTP dynamics, which has been
shown to be Markovian according to the trace-distance criterion \cite{Breuer2009}, but
\emph{not} according to the CP-divisibility criterion \cite{Rivas2010} in some parameter regimes \cite{Mazzola2010}.

Crucially, as the SL model yields dynamics that
correspond at each time instance to a phase covariant channel,
it can be specified employing the representation depicted in
\figref{fig:cov_maps} with:
\begin{eqnarray}
\label{eq:cov_pars_SLmodel}
\eta_{\perp}(t) & = & \frac{e^{-\frac{R}{2}\gamma t}}{f\!\left(\frac{R}{2},\gamma t\right)} = 1 - \frac{\gamma\gamma_0(2n+1)}{4} t^2 + o(t^2) ,\\
\eta_{\parallel}(t) & = & \frac{e^{-R\gamma t}}{f\!\left(R,\gamma t\right)} = 1 - \frac{\gamma\gamma_0(2n+1)}{2} t^2 + o(t^2), \nonumber \\
\kappa(t) & = & -\frac{1}{2n+1}\left(1-\frac{e^{-R\gamma t}}{f(R,\gamma t)}\right) = -\frac{\gamma \gamma_0}{2} t^2 + o(t^2). \nonumber
\end{eqnarray}
We have explicitly written down the short-time expansion of the noise parameters, as
in \eqnref{eq:coeffs_app}, whose short time-scaling is quadratic, $\beta_\perp=\beta_\parallel =\beta_\kappa=2$,
proving the SL model to exhibit the Zeno regime \cite{Facchi2008,*Pascazio2014}.
One may verify utilising \eqnsref{eq:see}{eq:soll} that the two pictures \eref{eq:gammas_SLmodel} and
\eref{eq:cov_pars_SLmodel} are indeed equivalent.

Consequently, using our derived general bound, \eqnref{eq:step2} in the main text, and the expansions \eref{eq:cov_pars_SLmodel},
we may write the predicted ultimate limit on achievable precision for the SL model as
\begin{equation}
\lim_{N \rightarrow \infty} \frac{\Delta^2 \omega_N\,T}{N^{-3/2}}
\geq \sqrt{2\gamma\gamma_0\max\{n,\frac{1}{4}\}}=\sqrt{2\gamma\gamma_0 n} = D_\t{SL},
\label{eq:bound_SL}
\end{equation}
assuming without loss of generality that $n\!>\!1/4$.

On the other hand, as the SL model leads to a non-unital phase covariant channel,
in order to calculate the \emph{exact} asymptotic constant for the GHZ-based strategy we
cannot simply utilise \eqnref{eq:dupa} to establish $\alpha_t$ in \eqnref{eq:Fghz_as}, but
must solve numerically \eqnref{eq:alphat_opt}. Yet, as $(2n+1)\gamma\gamma_0/2\!\le\!\alpha_t\!\le\!(n+1)\gamma\gamma_0$
according to \eqnref{eq:alphatrange} for the SL model, we may always set
$\alpha_t\!=\!(2n+1)\gamma\gamma_0/2$ (that we numerically verify to be suboptimal
up to negligible precision) to obtain
\begin{eqnarray}
\lim_{N \rightarrow \infty} \frac{\Delta^2 \omega^\t{\tiny GHZ}_N\, T}{N^{-3/2}}
&=&
\sqrt{\alpha_t \,\ee^{\frac{(2n+1)\gamma \gamma_0}{2 \alpha_t}}}  \cosh\!\left(\frac{\gamma  \gamma_0}{4 \alpha_t}\right) \\
&\lessapprox&
\ee^{\frac{n}{2n+1}}\frac{1+\ee^{\frac{1}{2n+1}}}{2\sqrt 2}\!\sqrt{(2n+1)\gamma\gamma_0} \nonumber \\
&=&
\sqrt{\frac{\ee}{2}\gamma\gamma_0\,2n\left[1+o(1/n)\right]} \nonumber\\
&\underset{n\gg1}{\approx}&
\sqrt{\frac{\ee}{2}}D_\t{SL}.
\label{eq:GHZ_SL}
\end{eqnarray}
Hence, the GHZ-based strategy for the SL model and a reservoir with $n\!\gg\!1$
saturates the asymptotic bound \eref{eq:bound_SL} up to a constant
factor $\sqrt{\ee/2}$.

\end{document}